\documentclass[nofootinbib,floatfix,aps,pra,twocolumn]{revtex4}
\usepackage{amsmath,epsf,dcolumn,natbib,graphicx}
\usepackage{xcolor}

\newcommand{\be}{\begin{equation}}
\newcommand{\ee}{\end{equation}}
\newcommand{\beq}{\begin{equation}}
\newcommand{\eeq}{\end{equation}}
\newcommand{\bea}{\begin{eqnarray}}
\newcommand{\eea}{\end{eqnarray}}

\begin{document}

\title{Comparison of Density Functional Approximations \\ 
and the Finite-temperature Hartree-Fock Approximation in Warm Dense Lithium}

\author{Valentin V.~Karasiev}
\email{vkarasev@qtp.ufl.edu}
\affiliation{Quantum Theory Project, 
Departments of Physics and of Chemistry, P.O. Box 118435, 
University of Florida, Gainesville FL 32611-8435}

\author{Travis Sjostrom}
\affiliation{Quantum Theory Project, 
Departments of Physics and of Chemistry, P.O. Box 118435, 
University of Florida, Gainesville FL 32611-8435}

\author{S.B.~Trickey}
\affiliation{Quantum Theory Project, 
Departments of Physics and of Chemistry, P.O. Box 118435, 
University of Florida, Gainesville FL 32611-8435}

\date{19 July 2012}

\begin{abstract}

We compare the behavior of the finite-temperature Hartree-Fock model
with that of thermal density functional theory using both ground-state
and temperature-dependent approximate exchange functionals.  The test system is
bcc Li in the temperature-density regime of warm dense matter (WDM).
In this exchange-only case, there are significant qualitative
differences in results from the three approaches.  Those differences may be
important for Born-Oppenheimer molecular dynamics studies of WDM with
ground-state approximate density functionals and thermal occupancies.
Such calculations require reliable regularized potentials over a 
demanding range of temperatures and
densities.  By comparison of  pseudopotential
and all-electron results at ${\mathrm T} = 0$K for small Li clusters of local bcc symmetry
and bond-lengths equivalent to high density bulk Li, we determine
the density ranges for which standard projector augmented wave (PAW) 
and norm-conserving pseudopotentials are reliable.  Then we construct
and use all-electron PAW
data sets with a small cutoff radius which are valid for lithium 
densities up to at least 80 g/cm$^3$. 

\end{abstract}


\maketitle

\renewcommand{\baselinestretch}{1.05}\rm

\section{Introduction}

Warm dense matter (WDM) encompasses the region between conventional condensed
matter and plasmas.  WDM occurs on the pathway to inertial confinement
fusion and is thought to play a significant role in the structure of
the interior of giant planets.  The theoretical and computational
description of WDM is important for understanding and performing
experiments in which WDM is created \cite{HEDLPreport.2009}.  Two
parameter ranges which are very different from those in standard
condensed matter physics characterize WDM: elevated temperature (from
one to a few tens of eV) and high pressure (up to thousands of
GPa). These ranges are challenging computationally because the
standard solid state physics methods become very expensive (due to
high temperature) or standard approximations used in those methods
cease to work (due to high material density). From the plasma side,
the temperature and pressure are not high enough to employ classical
approaches.

A combination of a quantum statistical mechanical description of the electrons, 
and classical molecular dynamics for ions is a standard 
theoretical and computational approach to WDM at present.   
Usually the quantum statistical
mechanics is handled via finite-temperature density functional theory (ftDFT) 
\cite{Mermin65,Stoitsov88,Dreizler89}.   
There is a substantial literature, too large to review here, 
about such calculations 
at zero temperature 
via Born-Oppenheimer 
molecular dynamics (BOMD) or Car-Parrinello MD, with 
DFT implemented via the Kohn-Sham (KS) procedure 
for the electronic degrees of freedom.  The pertinent point is that
the same techniques can be applied to the finite-temperature case  
\cite{Alavi94,Silvestrelli99,Surh01,Desjarlais02,Galli04,Mazevet04,Collins05,%
Mazevet05,Recoules06,Faussurier09,Horner09,Recoules09,Vinko09,Wunsch09,%
Clerouin10}.  The combination, called {\it ab initio} molecular dynamics (MD),
is computationally costly at high temperature (for a given density) 
because of the large number of partially occupied KS orbitals 
which must be taken into account.

The great majority of the reported finite-temperature {\it ab initio} 
MD calculations 
use zero-temperature exchange-correlation (XC)
functionals, $E_{\rm xc}$, with Fermi thermal occupations to 
construct the electron density.  
In such calculations, the only $\mathrm T$-dependence in the
XC contribution to the free energy 
${\mathcal F}_{\rm xc}$ is through 
the $\mathrm T$-dependence of the electron density:
\be
{\mathcal F}_{\rm xc}[n({\mathbf r},{\rm T}),{\rm T}] \approx E_{\rm xc}%
[n({\mathbf r},{\rm T})] \, ,
\label{FxcapproxExc}
\ee
with $n({\mathbf r},\rm T)$ the electron number density at temperature 
$\mathrm T$.  

Most ftDFT calculations with ground state XC functionals 
seem to have been done with the {\sc Vasp} 
\cite{VASP} or {\sc Abinit} \cite{AbInit} codes using either the 
local density approximation
(LDA) for $E_{\rm xc}$ \cite{VWN80,PZ81,Perdew.Wang.1992} or the 
Perdew-Burke-Ernzerhof generalized gradient approximation (GGA) 
functional \cite{PBE}.

The orbital-free density functional theory (OF-DFT) treatment of
electronic degrees of freedom is a less expensive alternative to 
orbital-dependent methods such as KS.  OF-DFT in principle provides the same
quantum-mechanical treatment of electrons as KS DFT,
but the lack of accurate orbital-free
approximations for the kinetic energy functionals has limited the use
OF-DFT, even at 
standard conditions. In contrast, the high density of the WDM regime is
favorable for use of the OF-DFT approach, which is a motive for
developing functionals.  The standard 
KS approach clearly must be used to test and calibrate such OF-DFT functionals.
The limitations and consequences of various
choices in those thermal KS calculations have not seen much 
detailed attention however. 
Two closely related sets of potentially significant issues occur. 

First, the use of ground-state functionals in a ftDFT 
calculation inevitably raises a topic for fundamental DFT, namely, the
adequacy, accuracy, and scope of Eq.\ (\ref{FxcapproxExc}). Relative
to the number of calculations, there are comparatively few studies to
assess this approach against others
\cite{Surh01,Desjarlais02,Faussurier09,Recoules09,%
  Vinko09,Wunsch09,Clerouin10,Danel06}.  Ref.\ \onlinecite{Surh01}
shows that the maximum density of the Al shock Hugoniot is increased
about 5\% or less by use of a temperature-dependent functional of the
Singwi-Tosi-Land-Sj\"olander (STLS) type \cite{Tanaka85}.  Ref.\  
\onlinecite{Danel06} made essentially the same comparison but with
respect to simple Slater exchange (in Hartree atomic units) 
\vspace*{-4pt}
\bea
E_{\rm x}[n] & = &  \int d{\mathbf r} n({\mathbf r}) %
\epsilon_{\rm x,S}[n({\mathbf r})]   \nonumber \\
\epsilon_{\rm x,S}[n({\mathbf r})] & := & C_{\rm x,S} n^{1/3}({\mathbf r})  %
\nonumber \\
C_{\rm x,S} &:=& -\frac{3}{4}\left( \frac{3}{\pi} \right)^{1/3}  \, .
\vspace*{-4pt}
\label{ExS}
\eea
and with the added complication [for the purpose of assessing 
Eq.\ (\ref{FxcapproxExc})] of use of an OF-DFT  
approximation.  Ref.\ \onlinecite{Desjarlais02} compared calculations for ground-state LDA and PW91 GGA \cite{PW91} functionals. 
Faussurier {\it et al.} \cite{Faussurier09} compared the electrical
conductivity of Al computed with the T-dependence from classical-map
hypernetted chain scheme \cite{PerrotDharmawardana00} versus 
ground-state LDA.  They concluded that the effects on conductivity 
are small in the 
WDM regime but become increasingly important as the energy density
increases.   W\"unsch {\it et al.}\cite{Wunsch09} reversed the
perspective and used ftDFT calculations with a ground-state XC 
functional to calibrate 
hypernetted chain approximations, hence assumed the validity 
of Eq.\ (\ref{FxcapproxExc}). Vinko {\it et al.} compared 
ground-state GGA calculations of free-free opacity for 
Al with an RPA model and
found semi-quantitative agreement at lower photon energies with 
increasing disagreement at higher ones, all over the range 
$0 \le {\mathrm T} \le 10$ eV \cite{Vinko09}.    As an aside,
we note that the same issues of use of ground-state approximate
XC functionals in a T-dependent context can arise in average-atom
models \cite{Feynman..Teller.1949,Liberman79,Rozsnyai72_91,Purgatorio,StarrettSaumon12}.
 
The second set of issues involves computational technique.  The
primary focus is control of the effects of
pseudopotentials (or regularization of the nuclear-electron 
interaction). These are ubiquitous in the highly refined codes 
in use for both WDM and ground state calculations.  Clear insight 
into the behavior and limitations of functionals requires that 
the regularized potentials not introduce artifacts of their own.
The challenge is to test those potentials against high-quality 
all-electron (AE) results over the appropriate density range.  

An obvious issue associated with pseuodpotentials is the effect 
of a finite core radius upon compressibility
(hence, equation of state). 
Ref. \cite{Mazevet.Lambert..2007} shows that a norm-conserving
pseudopotential for boron with the standard cutoff radius ($r_c=1.7$ Bohr)
is not transferable to the high material density regime. 
In that work, the authors built an \lq\lq all-electron" pseudopotential
with small $r_c=0.5$ Bohr and tested its transferability   
to very high material density by comparison with the Thomas-Fermi (TF)
limit calculated using an average-atom model \cite{Feynman..Teller.1949}.

Another issue is the extent to
which removal of core electrons has an unphysical effect on 
the distribution of ionization. A related issue is the effect 
that removing core levels has on Fermi-Dirac occupation numbers. At
fixed density, such 
core levels should be progressively depopulated with increasing
temperature.  Does the depopulation of pseudo-density levels behave
correctly?    A significant computational practice issue is
the minimum magnitude threshold for retention of occupation 
numbers.  That threshold is directly related to basis set
size or, equivalently, the plane-wave cutoff.   
We know of only one study of any of these questions
\cite{LevashovEtAl10}.  In it, all-electron calculations with the
full-potential linearized muffin-tin orbital methodology were used 
to benchmark projector augmented wave (PAW) calculations with a plane
wave basis.  Two metals, Al and W, were treated at $T \ne 0$K.  At 
least for W, it appears that different XC functionals were used for the
comparison.  Additionally, Ref.\ \onlinecite{LevashovEtAl10} used the
free-electron expression for the non-interacting electronic entropy, 
rather than the proper explicit dependence on occupation numbers $f_i$:
\be
{\mathcal S}_s = -k_B \sum_i \lbrace f_i \ln f_i + (1-f_i)\ln (1-f_i) \rbrace %
  \; .
\ee
Despite these differences, to the extent that their topics and ours
overlap, the findings are consistent.  

To establish a basis for comparison, first we consider the issues of
regularized potentials.  We consider both ordinary pseudopotentials (PPs)
and the pseudopotential-like PAW technique.
Those tests are against all-electron (bare Coulomb nuclei potential)
calculations for small Li clusters of bcc symmetry.  We establish a
PAW which demonstrably is reliable for the density range of interest.
Then we study the behavior and limits of the use of ground-state X
functionals in ftDFT by
comparison of finite-temperature Hartree-Fock (ftHF) and 
DFT exchange-{\it only} results.  For clarity of interpretation, 
all the bulk solid
calculations reported here were performed at fixed ionic positions
corresponding to an ideal bcc structure for Li. 

\section{Codes}

We used the {\sc atompaw} code \cite{Holzwarth..Matthews.2001} to form 
the PAWs. For periodic systems, we used three codes, {\sc Abinit} 
vers.\ 6.6 \cite{AbInit}, {\sc Vasp} vers.\ 5.2 \cite{VASP}, 
and {\sc Quantum-Espresso} ver.\ 4.3 \cite{QEspresso}.  
All three are plane-wave, PP codes.  All three also implement
PAWs.  {\sc Abinit} and {\sc Quantum-Espresso} are open source.
Technical details of the ftHF calculations are discussed below.     For the
all-electron calculations on finite clusters, we used conventional 
molecular gaussian basis techniques as embodied in 
the {\sc Gaussian 03} program \cite{Gaussian03}. 

\section{Regularized Potentials}

Diverse PP techniques commonly are used in KS
calculations to reduce computational cost by excluding the core
electrons from the self-consistent field (SCF) procedure and to
regularize the singular external potential in order to use 
an efficient, compact plane wave basis set.
Excluding core electrons implicitly invokes the frozen core
approximation ({\it i.e.}, the omission of core electrons from the SCF
procedure).  That approximation generally is well-justified in standard
conditions.  There, the core electrons are uninvolved in chemical
bonding and their state is essentially independent of the chemical
environment.  The validity of this justification is not obvious for 
the WDM regime.
In it, all electrons become important for
correct evaluation of the Fermi occupancy at high temperature and 
correct description of the electron density at high external pressure.
As a consequence, it is mandatory to include at least some core 
electrons in the solution of the relevant Euler equation (DFT or 
finite-temperature HF)
in the WDM regime. For light atoms this may mean an {\it all-electron} PP.
Those are, of course, a particular form of regularized potential.
 
Generation of PPs usually is characterized by cutoff (or pseudization) radii, 
$r_c$. Values of $r_c$ are a compromise between softness 
of the PP (for compactness of plane wave basis sets) and correct 
description of the one-electron orbitals close to the nucleus.
Standard PPs are developed for use under near-equilibrium 
condensed matter and molecular conditions, hence their transferability 
to the WDM regime needs to be explored. For example, 
commonly 
$r_c$ is assumed to be somewhat smaller than half the nearest-neighbor 
distance between atoms so that there is no core overlap. 
There is no guarantee that such equilibrium
prescriptions are satisfactory for WDM studies. 

\subsection{Basic PAW formalism}

PAW concepts are summarized in Ref.\ \onlinecite{Torrent..Xu.2010}.  
We outline the relevant points here.   
The PAW valence electron energy is comprised of a 
pseudo-energy evaluated using a smooth pseudo-density and pseudo-orbitals
plus atom-centered corrections. An energy correction centered on atom $a$ is 
evaluated using an augmentation sphere of radius $r_c^a$.  Within
each sphere, the correction
replaces the valence pseudo-energy of atom $a$, $\tilde E_v^a$, by the 
valence energy  $E_v^a$ generated from the valence part of the 
all-electron atomic density
\be
E_v=\tilde E_v + \sum_a \Big( E_v^a-\tilde E_v^a \Big)\,.
\vspace*{-4pt}
\label{E1}
\ee
Detailed descriptions of each term in Eq.\ (\ref{E1}) are given, for example, 
in  Ref.\ \cite{Torrent..Gonze.2008}.  Here the issue is
treatment of core density contributions to the XC energy, as discussed
in  that reference. In the scheme due to Bl\"ochl \cite{Bloechl.1994}, the XC 
energy is expressed as
\be
E_{\rm xc}=E_{\rm xc}[\tilde n + \tilde n_c] + 
\sum_a \Big(E_{\rm xc}[n^a + n_c^a] - E_{\rm xc}[\tilde n^a + \tilde n_c^a]%
 \Big)\,,
\label{E2}
\ee
where $n^a$ and $n_c^a$ are atom-centered valence and core electron 
charge densities corresponding to all-electron atomic orbitals, 
$\tilde n^a$ and $\tilde n_c^a$ are atom-centered 
valence and core electron pseudo-densities,
and $\tilde n$, $\tilde n_c$ are total valence and 
core electron pseudo-densities. 
The idea behind Eq.\ (\ref{E2}) is that the third term, which corresponds to
atom-centered contributions of pseudo-densities (evaluated within augmentation 
spheres, radii $r_c^a$), cancels the corresponding atom-centered pseudo-density
contributions (evaluated over all space) in the first term,  
and the canceled contribution is replaced by the second term,
which is evaluated with atom-centered all-electron densities 
(again within the augmentation spheres only).  

The Kresse scheme \cite{Kresse.Joubert.1999} introduces a 
valence compensation charge density, $\hat n$, as well. Its
purpose is to reproduce the multipole moments 
of the all-electron charge density outside the augmentation 
spheres \cite{Torrent..Gonze.2008}. For the XC contribution, $\hat n$ is 
added to 
the pseudo-densities in the functionals in Eq. (\ref{E2}) to give
\be
E_{\rm xc}=E_{\rm xc}[\tilde n + \tilde n_c+\hat n] + 
\sum_a \Big(E_{\rm xc}[n^a + n_c^a] - 
E_{\rm xc}[\tilde n^a + \tilde n_c^a + \hat n^a] \Big)\,.
\label{E3}
\ee
This procedure can cause problems with GGA XC functionals; see 
Ref.\ \onlinecite{Torrent..Xu.2010}. 

There are what are called all-electron PAWs, which in
essence are regularized potentials for all-electron calculations. 
In customary notation, an ``N-electron'' PAW retains N 
electrons in the valence.  Thus a 3-electron (``$3e^-$'') PAW 
calculation for Li is an all-electron, regularized-potential calculation.  

\subsection{PAW and high density lithium}
\label{PAWhighdensLi}

We tested the PAW approach by calculating the pressure of bcc Li 
over a large range of  material  densities, from approximate
equilibrium, $\rho_{\rm Li}=0.5$ g/cm$^3$,  to $\rho_{\rm Li}=25.0$ g/cm$^3$,
all at $\mathrm{T} = 100$K.  (The equilibrium density from simple 
Slater LDA all-electron calculations  is 0.54 g/cm$^3$, or lattice constant 
6.59 Bohr, close to the experimental value; 
see Ref.\ \onlinecite{BoettgerTrickey85}.  Newer LDAs give somewhat
contracted results; see below.)
Three different PAW data sets were used for each LDA and GGA 
exchange-correlation functional: 
(i) the standard set with compensation charge density 
included from Ref. \onlinecite{atompaw.Li}, (ii) a set with the 
same cutoff radius  ($r_c=1.61$ Bohr) but without compensation
charge density, and (iii) a set we generated with $r_c=0.80$ Bohr and no 
compensation charge density. 
The Perdew-Wang (PW) and Perdew-Zunger (PZ) LDAs 
\cite{Perdew.Wang.1992,PZ81} and Perdew-Burke-Ernzerhof GGA
\cite{Perdew..Ernzerhof.1996} (PBE) XC functionals
were used. 

The upper segment of Table \ref{tab:PAW-a0} compares
the calculated bcc Li equilibrium lattice constants and bulk moduli 
for the various combinations. These were done with {\sc Abinit}.  
The lattice constant and bulk modulus were 
obtained by fitting the calculated total energies per cell to 
the stabilized jellium model equation of state (SJEOS) form \cite{SJEOS.2001}.
One sees that the exclusion of
the compensation density slightly decreases the lattice constant for
both PW and PBE functionals.  The results are essentially unchanged
when the $r_c$ value is decreased to 0.80 Bohr.  

Table
\ref{tab:PAW-a0} also summarizes results obtained using both 
{\sc Quantum-Espresso} and {\sc Vasp}.  The lattice constant 
and bulk modulus again were obtained via fitting to the SJEOS 
form in all cases. The results for 
{\sc Vasp} come from using the 
PAW pseudopotentials supplied with the code itself. There is excellent 
agreement between {\sc Quantum-Espresso} and {\sc Abinit}
results when the same $3e^-$ PAW data set is used. 
The {\sc Vasp} PBE $3e^-$ results do not agree as well,
consistent with the findings of Ref.\ \onlinecite{Torrent..Xu.2010} regarding
the effects of the valence compensation charge density contribution. 
Two LDA PPs also were used with {\sc Quantum-Espresso},
namely the $1e^-$ Von Barth-Car and 
$3e^-$ norm-conserving pseudopotentials (both taken from 
the {\sc Quantum-Espresso} web page).
The lattice constant corresponding to the first of these PPs is 
underestimated as compared to other PZ LDA calculations,
independent confirmation of the importance of the $3e^-$ treatment.
For the $1e^-$ (Vanderbilt ultrasoft) and $3e^-$ (norm-conserving) 
PBE PPs
(again taken from the {\sc Quantum-Espresso} web page), 
the lattice constant is slightly overestimated and 
the bulk modulus is underestimated 
by the $1e^-$ pseudopotential. The $3e^-$ results are in 
nearly perfect agreement with the PAW data.

\begin{figure}
\begin{tabular}{c}
\hspace{-4.5cm} \epsfxsize=4.1cm \epsffile{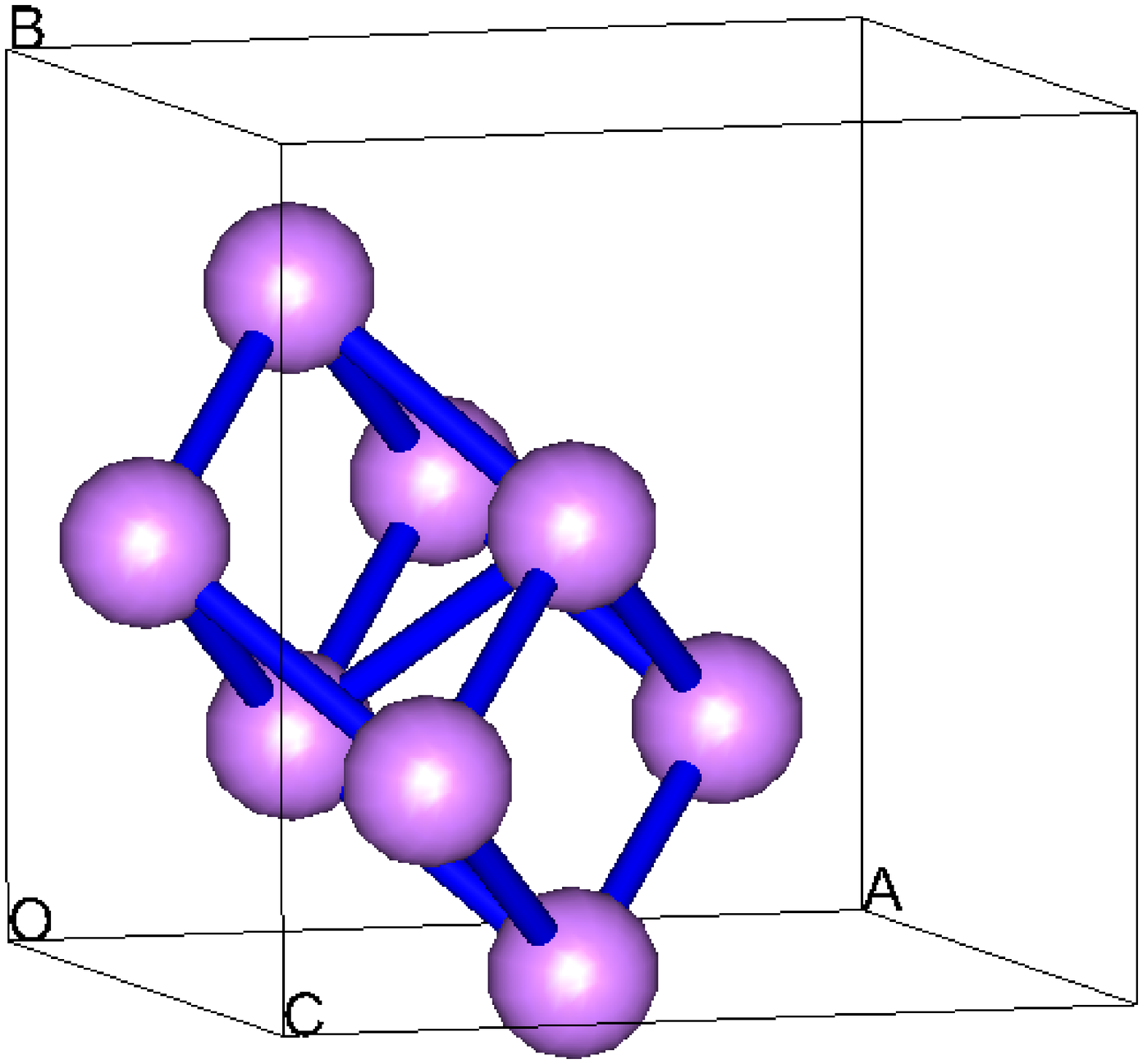}
\end{tabular}
\vskip -3.3cm
\begin{tabular}{c}
\hspace{4.0cm} \epsfxsize=4.1cm \epsffile{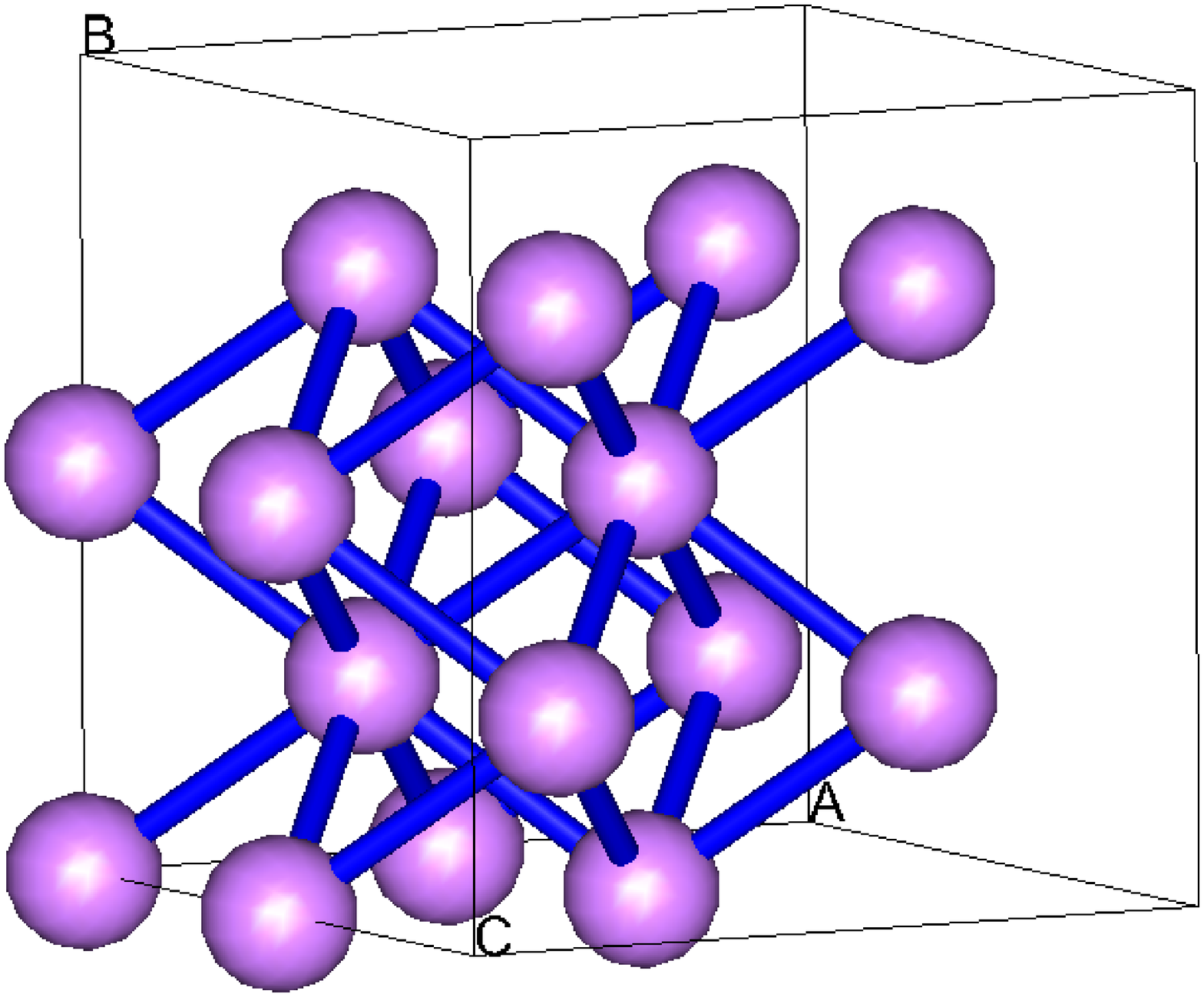}
\end{tabular}
\caption{
Bcc Li$_8$ (left panel) and Li$_{16}$ (right panel) clusters 
used to test PAW calculations. 
}
\label{bcc-Li-clusters}
\end{figure}

To assess the PAW method for high material density,  we compared PAW 
and true all-electron (bare Coulomb potential) 
results for two small lithium clusters with local bcc symmetry; see  
Fig. \ref{bcc-Li-clusters}. The interatomic distances in both 
clusters were set equal to the nearest-neighbor distance in bulk bcc-Li for 
densities in the range $0.5 \le \rho_{\rm Li} \le 150$ g/cm$^3$.
The all-electron calculations were done with the {\sc Gaussian 03} code
and two basis sets, 6-311++G(3df,3pd) and cc-pVTZ.  For the LDA 
calculations, we used the Vosko, Wilk, and Nusair parameterization 
(VWN) \cite{VWN80}; it is
very close to the PZ parameterization and based on the same data. 
For the GGA functional we used PBE. 
For high densities, $\rho_{\rm Li} \ge 50$ g/cm$^3$, we did 
additional calculations
with cc-pV5Z (8-atom cluster) and cc-pVQZ (16-atom cluster) basis sets.
The PAW calculations
were done with the {\sc Abinit} code.  In it, the clusters were centered in 
a large cubic super-cell of size $L$.  For the standard PAW, we used 
$L=15$ $\AA$ with an energy cutoff  1000 eV, while  $L=12$ $\AA$ with 
energy cutoff 3000 eV was used for the small $r_c$ PAW.
The  PZ LDA and PBE GGA functionals were used in these calculations. 
Note that the difference in behavior  between 
PZ and PW is essentially negligible for the purposes of this study. 

\begin{figure}
\includegraphics[angle=-00,width=7.2cm]{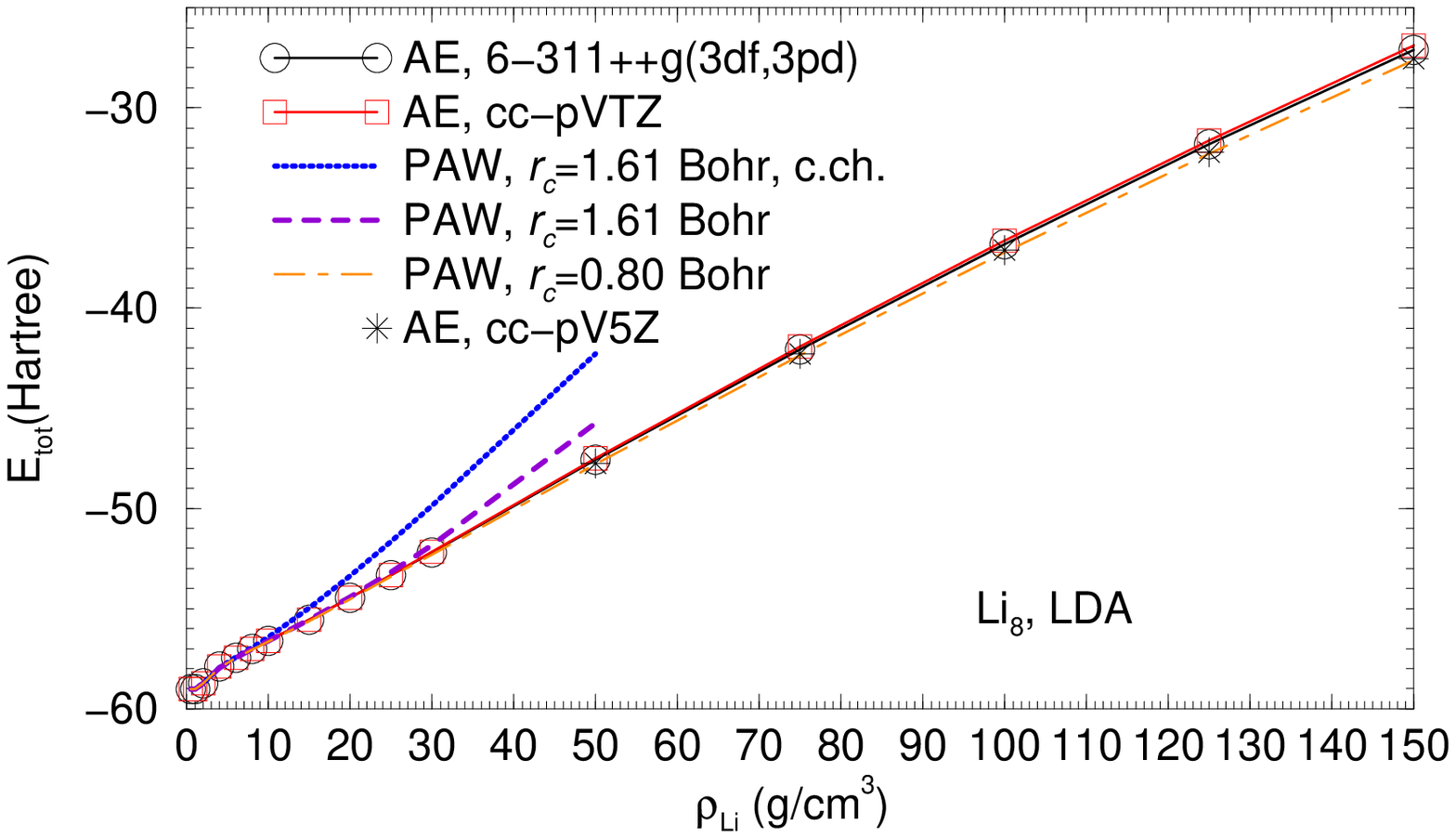}
\includegraphics[angle=-00,width=7.2cm]{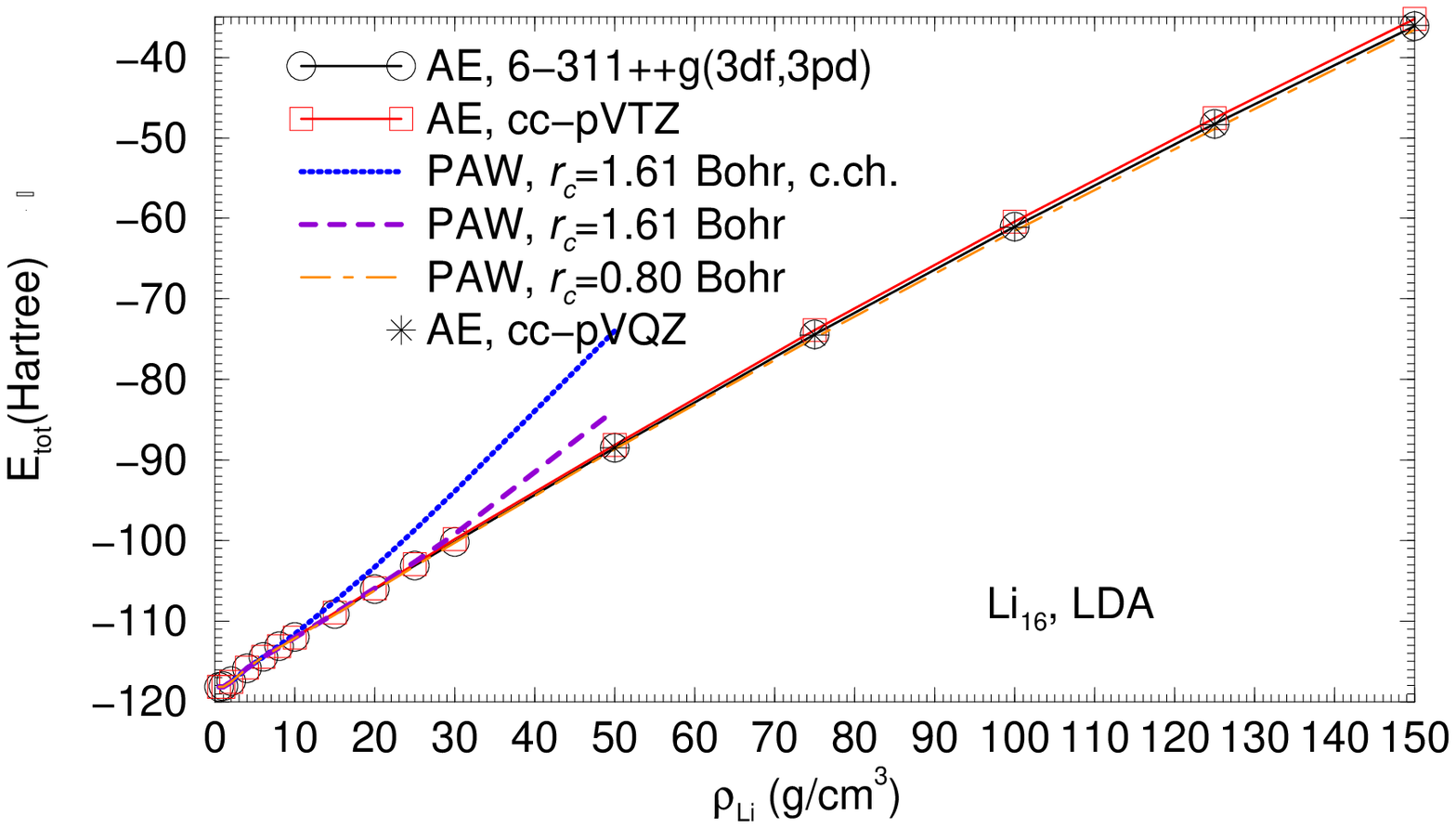}
\caption{
All-electron (VWN correlation) and PAW (PZ correlation) LDA total 
energies for the Li$_8$ and Li$_{16}$  clusters.
}
\label{E-R.clusters.LSDA}
\end{figure}

\begin{figure}
\includegraphics[angle=-00,width=7.2cm]{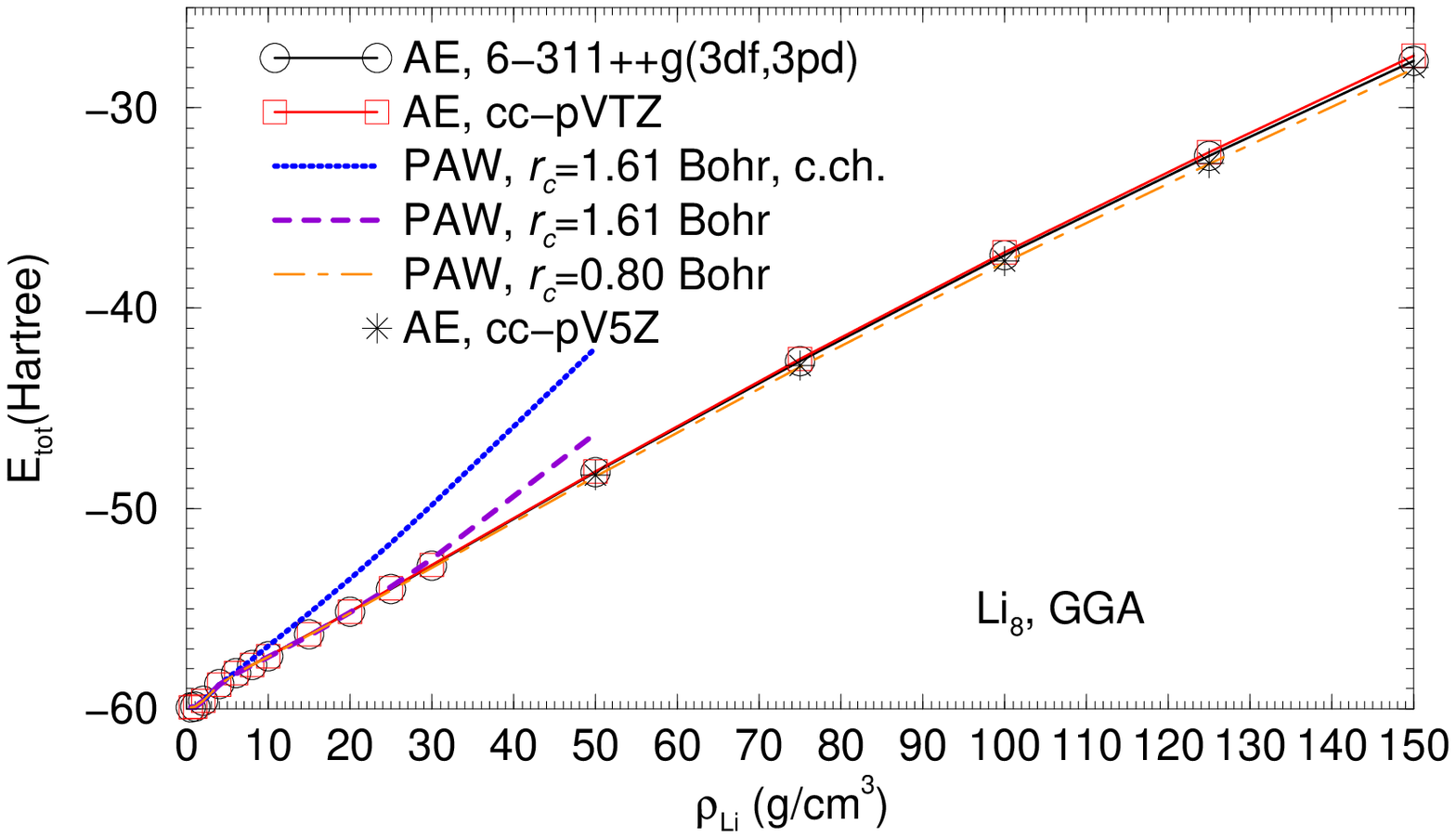}
\includegraphics[angle=-00,width=7.2cm]{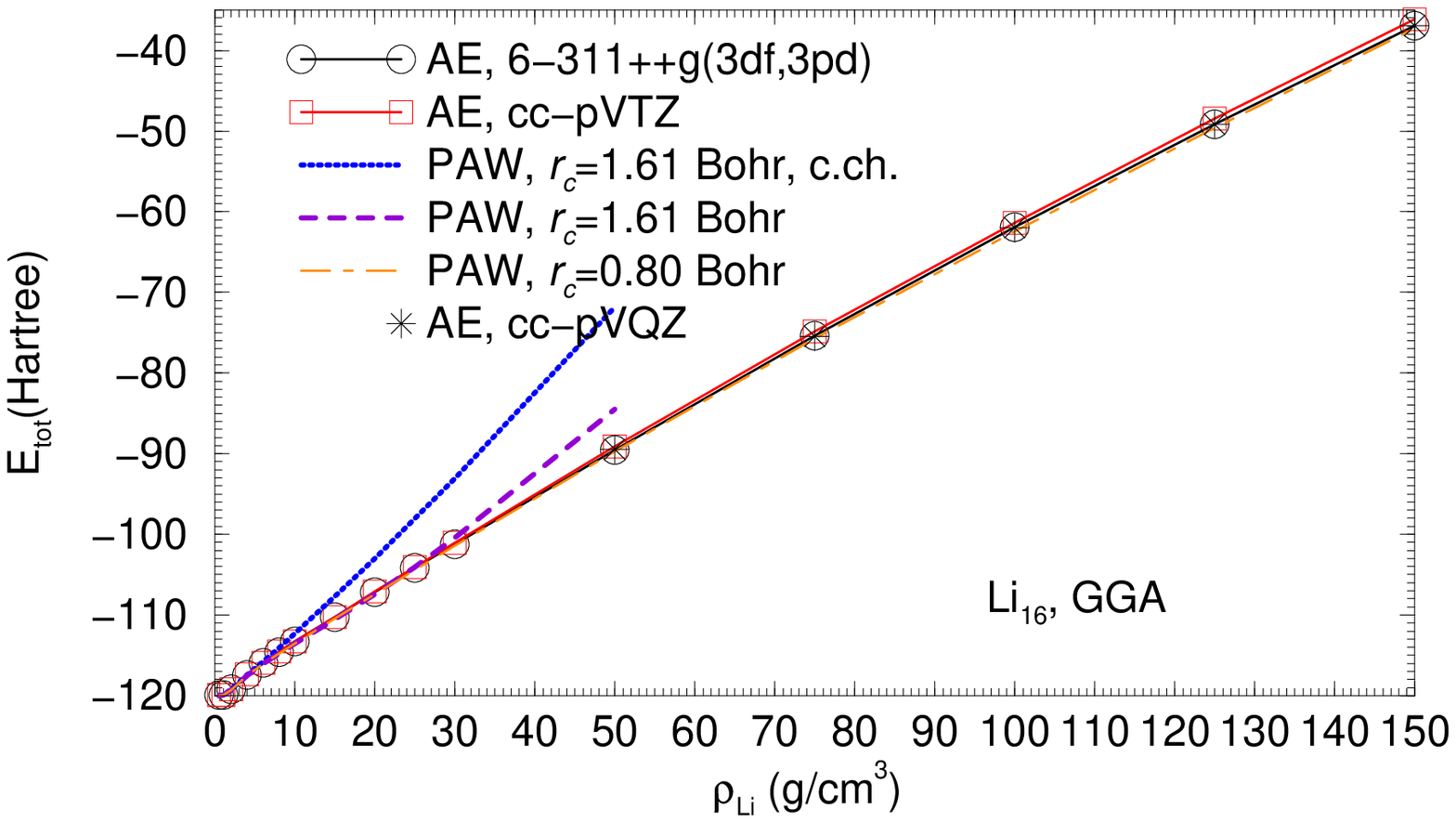}
\caption{
All-electron and PAW GGA total energies for the Li$_8$ and Li$_{16}$ 
clusters. 
}
\label{E-R.clusters.PBE}
\end{figure}

Fig.\ \ref{E-R.clusters.LSDA} shows all-electron and PAW LDA 
total energies for the two clusters as a function of distance 
corresponding to the stated bulk density. 
Fig.\ \ref{E-R.clusters.PBE} shows the corresponding GGA results.  
The behavior of the two clusters is quite similar.
For the standard PAW data set (labeled ``(i)'' previously), 
the total energy starts to deviate from the all-electron (AE) values  
at approximately $\rho_{\rm  Li}^{clust-crit1}=8.0$ g/cm$^3$. 
For the standard PAW set without
compensation density [set (ii)], the critical 
density $\rho_{\rm  Li}^{clust-crit2}$  is approximately 25
g/cm$^3$.  In contrast, the PAW with small $r_c$ and no compensation 
density [set (iii)] gives essentially perfect agreement with the AE
results for the whole density range. For densities up  to 30 g/cm$^3$, 
two basis sets, 6-311++G(3df,3pd) and cc-pVTZ, give essentially 
the same quality results.  
At high density (50 g/cm$^3$ and up), the cc-pVTZ basis set 
energies lie above
the values corresponding to the 6-311++G(3df,3pd) basis. For those high 
densities, AE calculations done with the larger cc-pV5Z 
(8-atom cluster) and cc-pVQZ (16-atom cluster) basis sets
lower the total energy to the 6-311++G(3df,3pd) level (16-atom cluster) 
or slightly lower (8-atom cluster).  Once again there is essential
perfect agreement with the set (iii) PAW plane wave results.

The corresponding PBE GGA comparison of PAW and AE results, 
Fig.\ \ref{E-R.clusters.PBE}, shows that the critical densities 
for each PAW data set are almost identical for the 8-atom and 16-atom 
clusters. For the PAW data set (i), 
$\rho_{\rm Li}^{clust-crit1}\approx 6.0$ g/cm$^3$ is slightly lower than 
for the LDA case. For PAW data set (ii),
the critical density is essentially 
the same as for LDA ($\approx 25$ g/cm$^3$).
Once again, the small $r_c$ PAW data set (iii) gives good  agreement with
the AE results up to the maximum density considered (150 g/cm$^3$). 
We conclude that PAW data set (iii),  namely 
$r_c=0.80$ Bohr and no compensation charge, may serve for making  
reference KS calculations in the high density regime.  

\begin{table*}
\caption{\label{tab:PAW-a0}
Equilibrium lattice constant for bcc-Li, $a$ (Bohr) 
and bulk modulus, $B$ (GPa).
}
\begin{ruledtabular}
\begin{tabular}{llcccccccc}
&&\multicolumn{4}{c} {LDA} & \multicolumn{4}{c} {$\rm GGA$}\\
\cline{3-6}\cline {7-8}
Method & $r_c$ & \multicolumn{2}{c} {PW} & \multicolumn{2}{c} {PZ}& \multicolumn{2}{c} {PBE} \\ 
\cline{3-4}\cline {5-6} \cline{7-8} 
&& $a$ & $B$ & $a$ & $B$ & $a$ & $B$ \\ 
\hline
\hspace*{10pt}\\ [-8pt]
{\sc Abinit} ($3e^-$, PAW, c.ch.\tablenotemark[1])            & 1.61 & 6.363 & 15.1 & --    & --   & 6.513 & 14.0 \\ 
{\sc Abinit} ($3e^-$, PAW)                   & 1.61 & 6.363 & 15.0 & 6.360 & 15.1 & 6.497 & 13.8 \\ 
{\sc Abinit} ($3e^-$, PAW)                   & 0.80 & 6.362 & 15.1 & 6.359 & 15.1 & 6.497 & 13.9 \\ 
\hline
{\sc Q-Espreso} ($3e^-$, PAW, c.ch.\tablenotemark[1])         & 1.61 & 6.364 & 15.0 & --    & --   & 6.516 & 14.1 \\ 
{\sc Q-Espreso} ($3e^-$, PAW, c.ch.\tablenotemark[1])         & 0.80 & 6.362 & 15.0 & --    & --   & 6.497 & 13.8 \\ 
\hline
{\sc Q-Espreso} ($1e^-$)\tablenotemark[2]    &  -- & --    & --   & 6.320 & 15.1 & 6.724 & 11.9 \\ 
{\sc Q-Espreso} ($3e^-$)\tablenotemark[3]    & --   & --    & --  & 6.362 & 15.1 & 6.500 & 13.8 \\ 
\hline
{\sc Vasp} ($1e^-$, PAW, c.ch.\tablenotemark[1])              & 2.05   & --    & --   & 6.373 & 15.0 & 6.514 & 13.7 \\ 
{\sc Vasp} ($3e^-$, PAW, c.ch.\tablenotemark[1])            & 1.55-2.00& --    & --   & 6.362 & 14.9 & 6.505 & 13.8 \\ 
\end{tabular}
\tablenotetext[1]{Compensation charge density is included.}
\tablenotetext[2]{LDA: PZ exchange-correlation, nonlinear core-correction Von Barth-Car; 
GGA: PBE exchange-correlation, nonlinear core-correction, Vanderbilt ultrasoft pseudopotentials.}
\tablenotetext[3]{PZ and PBE semicore state $s$ in valence Troullier-Martins pseudopotentials.}
\end{ruledtabular}
\end{table*}

\begin{figure}
\includegraphics[angle=-00,height=3.4cm]{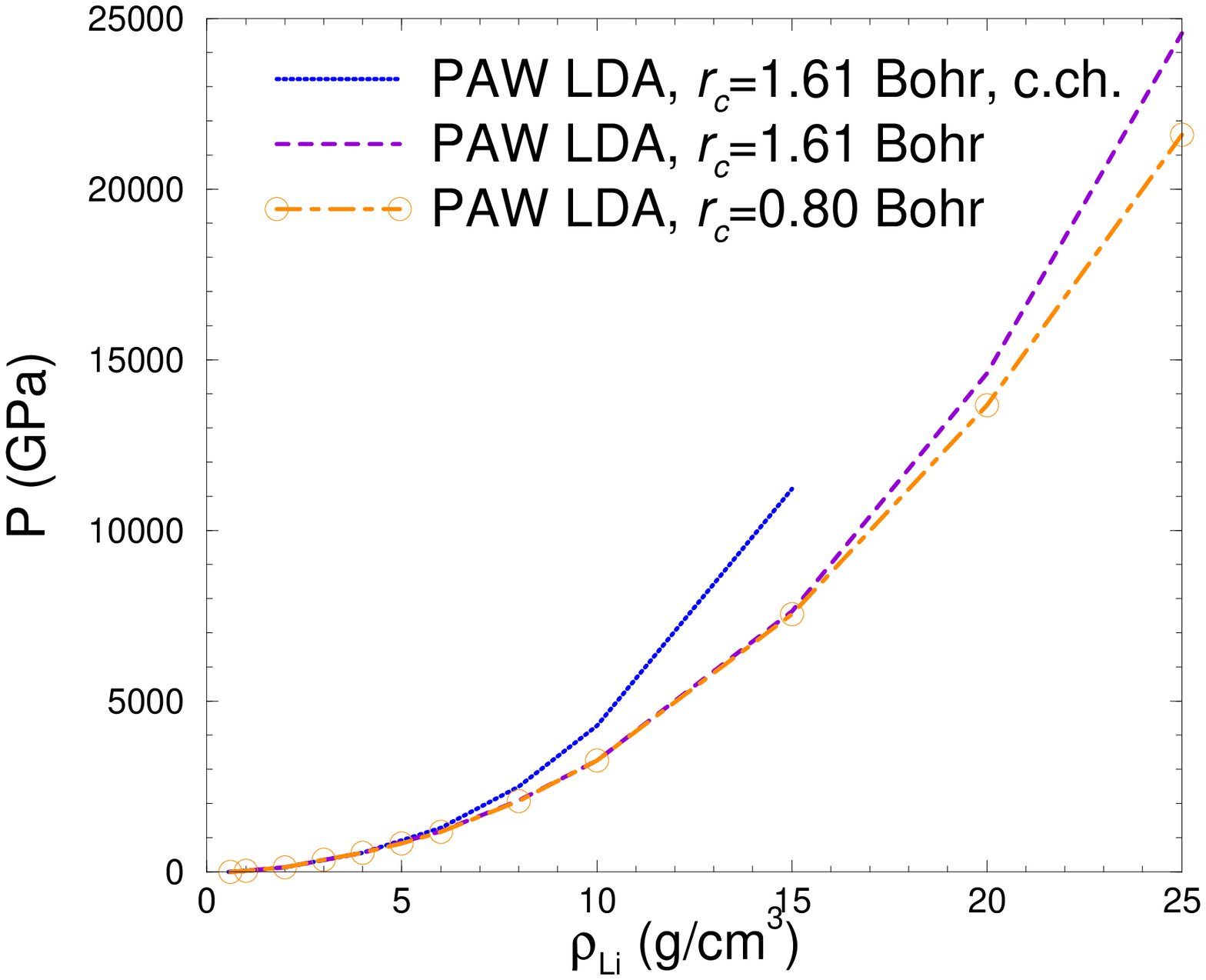}
\hspace{1pt} 
\includegraphics[angle=-00,height=3.4cm]{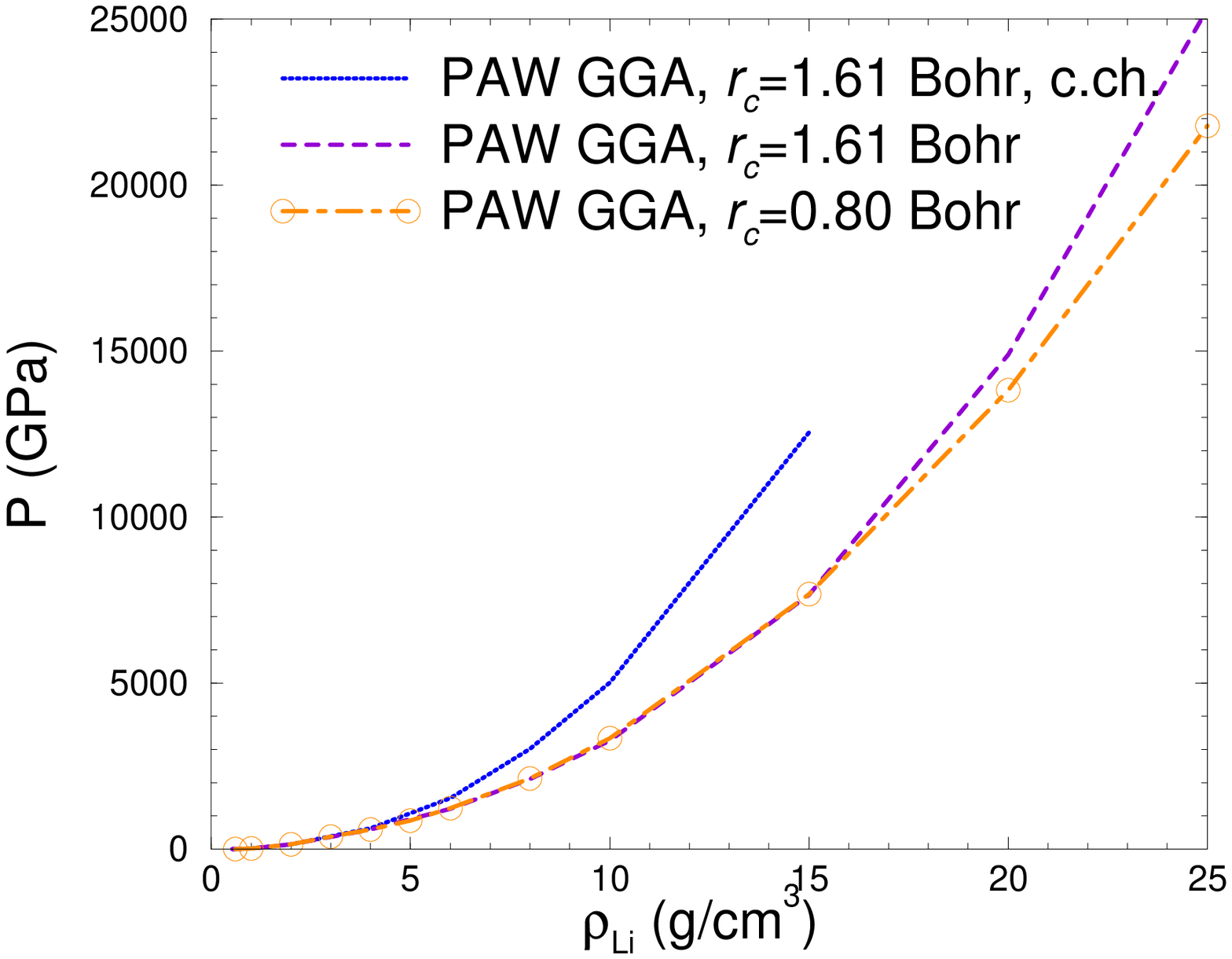}
\caption{
Pressure vs. material density from PAW LDA (PW correlation) (left panel) and
from PAW GGA (PBE XC) (right panel) calculations for bulk bcc-Li.
}
\label{P-R.bulk.PAW}
\end{figure}

The remaining validation issue is the effect of PAW or PP on the calculated
pressure.   A study \cite{Horner09}  of the EOS for warm, dense LiH found that 
the $3e^-$ PAW for Li in VASP calculations was necessary for T = 2, 4, and
6 eV and densities twice that of ambient and greater.    
Fig.\ \ref{P-R.bulk.PAW}  shows the bulk bcc Li pressure as a function of 
material density at T=100 K 
calculated using {\sc Abinit} with the same three PAW data sets as before for 
both the PW LDA and PBE XC functionals. (Use of the PZ LDA functional 
gives results 
indistinguishable from those from PW LDA on the scale of the figure.) 
One sees that the standard PAW data set (i) 
starts to overestimate the pressure 
at $\rho_{\rm Li}^{bulk-crit1}=6.0$ g/cm$^3$ for LDA and at a slightly
lower value for PBE.  PAW data set (ii) 
produces results which agree with the reference calculations 
({\it i.e.}, those from PAW set (iii)) for densities up to 
$\rho_{\rm Li}^{bulk-crit2}=15.0$ g/cm$^3$.
Comparison of critical density values in the clusters and in bulk 
shows that  $\rho_{\rm Li}^{bulk-crit1}$ is slightly lower
than $\rho_{\rm Li}^{clust-crit1}$.  A crude linear extrapolation
of the results from PAW data sets (i) and (ii) gives   
an estimated lower bound for the critical bulk density 
for the reference PAW data set  
$\rho_{\rm Li}^{bulk-crit3}$ to be 80 g/cm$^3$. Additional tests  
would be needed to get the actual value 
of $\rho_{\rm Li}^{bulk-crit3}$.  Such a determination is not required
for the present purposes.  

We observe that for fcc Al at $\mathrm T$=0 K, 
Levashov {\it et al.}\cite{LevashovEtAl10}  found that the 
standard VASP PAW pressures began to
deviate materially from all-electron values at about a compression
of seven. Since it was standard VASP, presumably that PAW included 
charge compensation, hence their 
result should correspond to our set (i) Li results, those labeled 
``PAW, $r_c$ = 1.61 bohr, c.ch.'' in Fig.\ \ref{P-R.bulk.PAW}.  It is clear 
that the deviation they found in fcc Al is at similar but modestly lower 
compression than we find for bcc Li.  

\begin{figure}
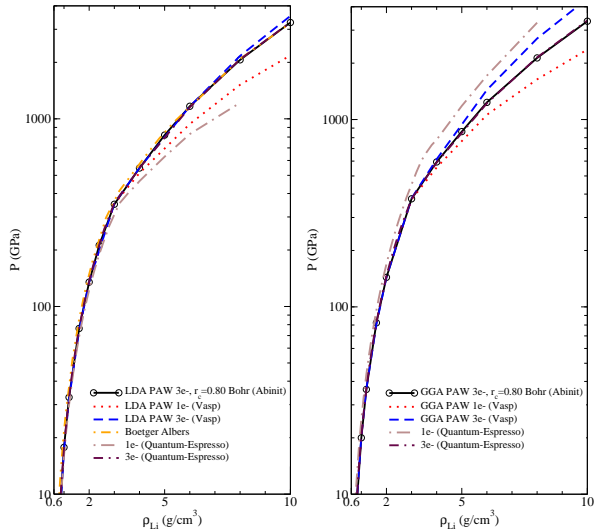

\includegraphics[angle=-00,height=7.0cm]{P-vs-R.LDA-PZ-0.80.LDA-Vasp.LDA-Siesta.Boettger-Albers.R0.6-10.log.eps}
\includegraphics[angle=-00,height=7.0cm]{P-vs-R.PBE-0.80.PBE-Vasp.PBE-Siesta.R0.6-10.log.eps}
\caption{
Validation of {\sc Vasp} 1$e^-$, 3$e^-$ PAW, 
{\sc Quantum-Espresso}  1$e^-$, 3$e^-$ PZ LDA (left panel) and 
{\sc Vasp} 1$e^-$, 3$e^-$ PAW,
{\sc Quantum-Espresso}  1$e^-$, 3$e^-$ PBE GGA (right panel) 
pseudopotential calculations: pressure as a function of density for bcc-Li calculated
at T=100 K.
}
\label{P-R.0.1kK.LSDA-abinit-vasp-siesta.R0.6-1.8}
\end{figure}



Figure \ref{P-R.0.1kK.LSDA-abinit-vasp-siesta.R0.6-1.8} compares  
pressure versus material density (T=100 K) for LDA (left panel)
and PBE GGA (right panel) for material densities in the range 
$0.6-10.0$ g/cm$^3$ obtained from 
{\sc Vasp} and {\sc Quantum-Espresso} using standard PPs
(the PAW provided with the {\sc Vasp} package and the norm-conserving PP 
taken from the {\sc Quantum-Espresso} 
web page), and 
reference results obtained with our PAW data set (iii) 
(for both the PZ LDA and  PBE GGA XC functionals.)  For LDA, we also 
show the earlier all-electron
results by Boettger and Albers \cite{Boettger.Albers.1989}. 
Observe first that our designation of the PAW (iii)
as a reference is substantiated by the
agreement with the all-electron LDA  calculation.  
The {\sc Vasp} $1e^-$ PAW LDA results start to deviate from the reference
values at about 3.0 g/cm$^3$.  However the $3e^-$ PAW LDA pressure
from {\sc Vasp}  
is in good agreement for densities up to 6.0 g/cm$^3$. 
{\sc Quantum-Espresso} results calculated with the $1e^-$ PZ LDA 
pseudopotential start to deviate from the reference results 
for density between 2 and 3 g/cm$^3$, whereas the 
$3e^-$ potential produces results which agree virtually perfectly  
for the full density range. For the 
GGA  case, the right-hand panel of 
Fig. \ref{P-R.0.1kK.LSDA-abinit-vasp-siesta.R0.6-1.8} 
shows that the situation is very similar except that 
both the $1e^-$ and $3e^-$ PAW {\sc Vasp} calculations
start to deviate from the reference results at almost the same density 
($\approx 3$ g/cm$^3$). 
{\sc Quantum-Espresso} $1e^-$ calculations 
overestimate the 
pressure for $\rho_{\rm Li}>0.8$ g/cm$^3$.  However, the 
{\sc Quantum-Espresso} $3e^-$ results are in 
virtually perfect agreement
with the reference PAW results for the whole range of densities.

\section{Finite temperatures}

\subsection{Pseudopotentials and level populations}
\label{FinTempPPlevel}

In finite-temperature calculations (either KS or HF), there is 
non-zero occupation of one-electron
levels which correspond to empty levels at $\mathrm{T} =0$ K (virtual states
or simply ``virtuals''). 
Satisfaction of some computational threshold for the smallest non-negligible
occupation number requires an increasingly large set of those virtuals to
be considered with increasing $\mathrm T$.  Concurrently there is  
depopulation of levels fully occupied at $\mathrm{T} = 0$ K.  
One would hope that PP methods which treat all
electrons self-consistently would be applicable for such finite-$\mathrm T$ 
calculations. A related issue is the validity of using
PPs which remove some of the core. A rough estimate 
of the relevant scale comes from taking the  
$1s$ ionization potential for the Li atom to be approximately the 
magnitude of the LDA Kohn-Sham $1s$ eigenvalue, about 51 eV.  Then PP  
treatment of Li $1s$ electrons as core might be expected to be 
applicable for temperatures much smaller than 51 eV. 
The question is the validity of any estimate of this sort, in particular,
how much smaller?   We remark that Levashov {\it et al.} \cite{LevashovEtAl10}
found that for ambient density Al, the PAW pressure deviated from the 
all-electron value at about $\mathrm T$ = 5-6 eV.  This is less
than  10\% of the magnitude of the LSDA 2p atomic KS eigenvalue (about 70 eV).

\begin{figure}
\begin{tabular}{c}
\hspace{-4.5cm} \epsfxsize=4.2cm \epsffile{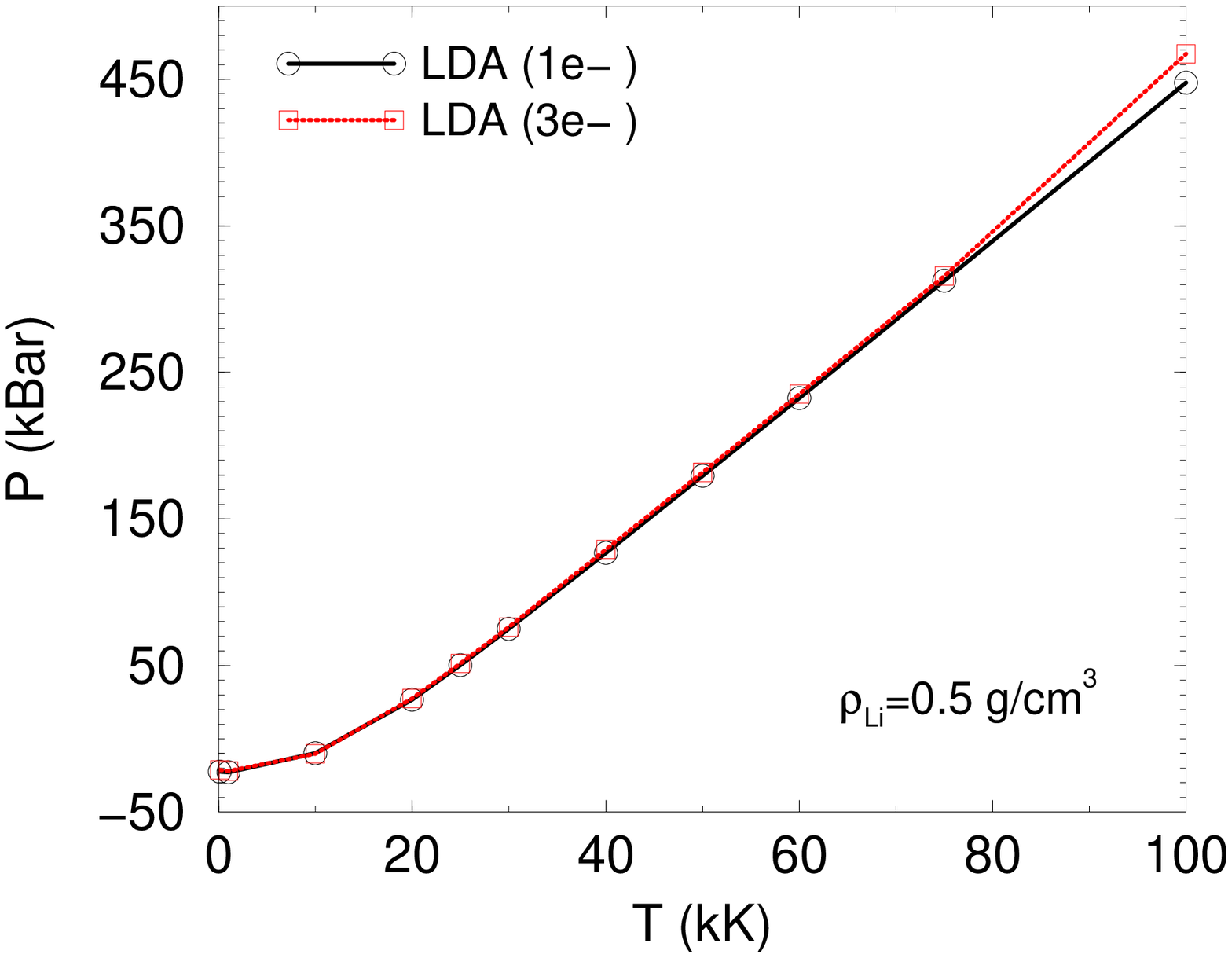}
\end{tabular}
\vskip -3.5cm
\begin{tabular}{c}
\hspace{4.0cm} \epsfxsize=4.2cm \epsffile{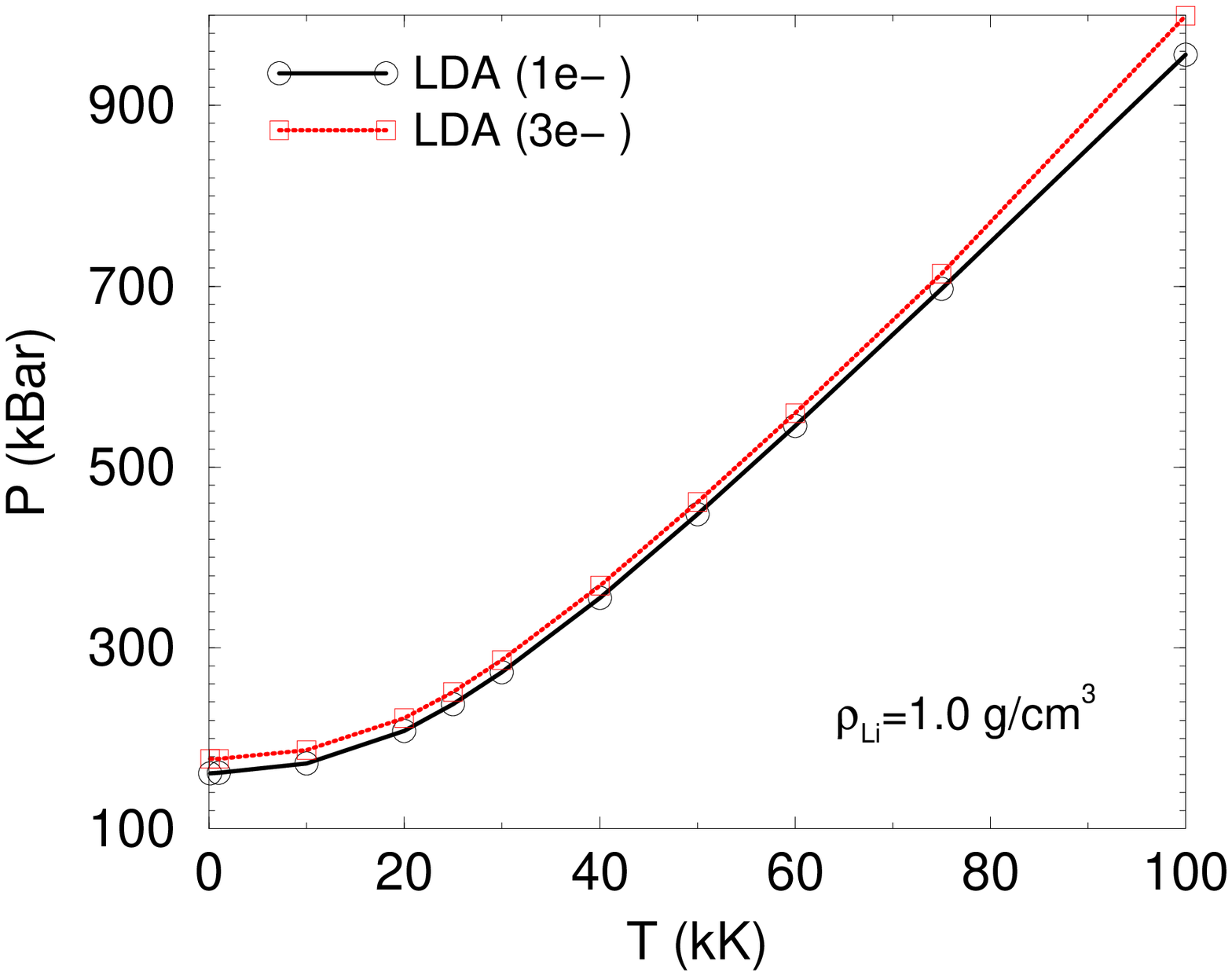}
\end{tabular}
\caption{
Comparison of pressure vs.\ temperature for bcc-Li obtained
with 3${\it e-}$ and 1${\it e-}$ pseudopotentials for the Perdew-Zunger LDA
exchange-correlation functional as 
implemented in {\sc Quantum-Espresso} (2-atom unit cell, 9x9x9 k-mesh, 128 bands).
Left panel: $\rho_{\rm Li}=0.5$ g/cm$^3$.
Right panel: $\rho_{\rm Li}=1.0$ g/cm$^3$.
}
\label{P-T.LDA-TM-1e-3e.QEspresso}
\end{figure}

First consider the comparative performance of the PPs.  
Figure \ref{P-T.LDA-TM-1e-3e.QEspresso} shows the hydrostatic pressure
as a function of temperature calculated using $1e^-$ and 
$3e^-$ norm-conserving pseudopotentials for the bcc-Li structure (fixed 
nuclear positions, $\rho_{\rm Li}=0.5$ and $1.0$ g/cm$^3$). 
Notice that this calculation uses a ground-state XC functional: there 
is no explicit temperature dependence in the PZ LDA XC functional. 
If the number of bands taken into account for a two-atom unit
cell for $\rho_{\rm Li}=0.5$ g/cm$^{-3}$ is 128, the occupation 
number of the highest energy bands is of the 
order of $10^{-6}-10^{-7}$.  Observe that the results from 
the $1e^-$ PP are in almost perfect agreement with those from
the $3e^-$ calculations for $\mathrm T$ up to 75,000 K, with small 
disagreement appearing at higher temperatures.  For low to moderate
compression, it appears that
the range of applicability of standard $1e^-$ 
norm-conserving pseudopotentials is at least up to ${\mathrm T} = $ 100 kK 
or about
8 -- 9 eV. This fits the rough argument based on the Li $1s$ KS eigenvalue,
with the criterion for ``much smaller'' being of order 20 \% at most.

Next comes the matter of significant fractional
occupation of ever-higher energy orbitals with increasing
temperature. A related issue is energy level shifting and reordering with
increased density, an effect known for ${\mathrm T} = 0$K Li 
\cite{BoettgerTrickey85,ZittelEtAl85}.  

\begin{figure}
\includegraphics[angle=-90,width=4.2cm]{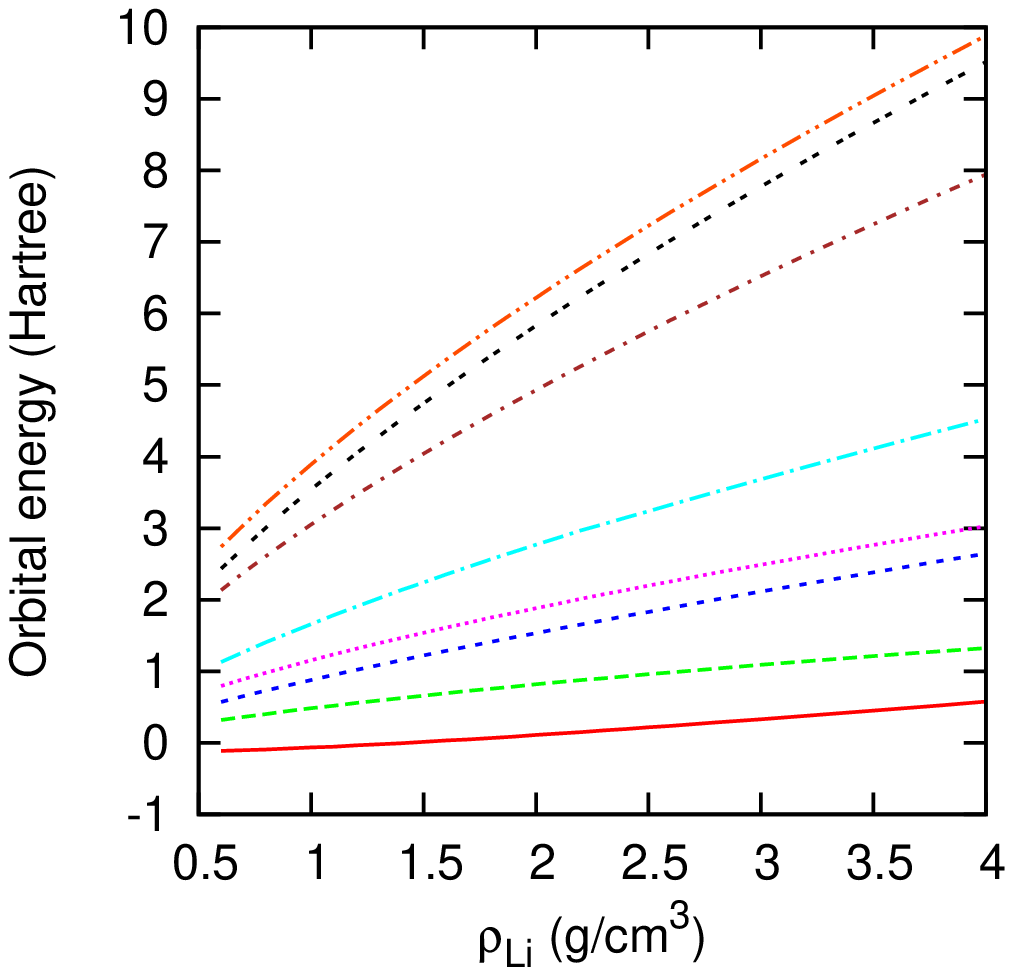}\\
\includegraphics[angle=-90,width=4.2cm]{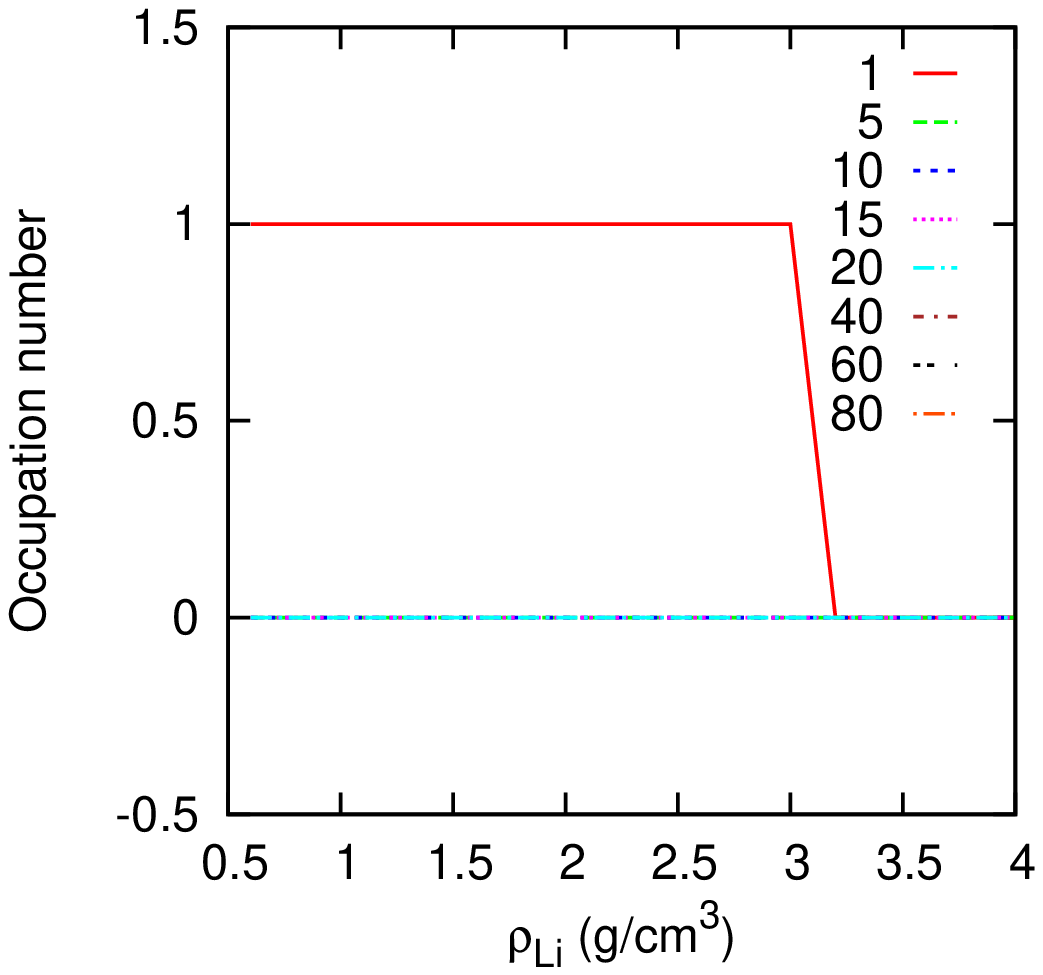}
\includegraphics[angle=-90,width=4.2cm]{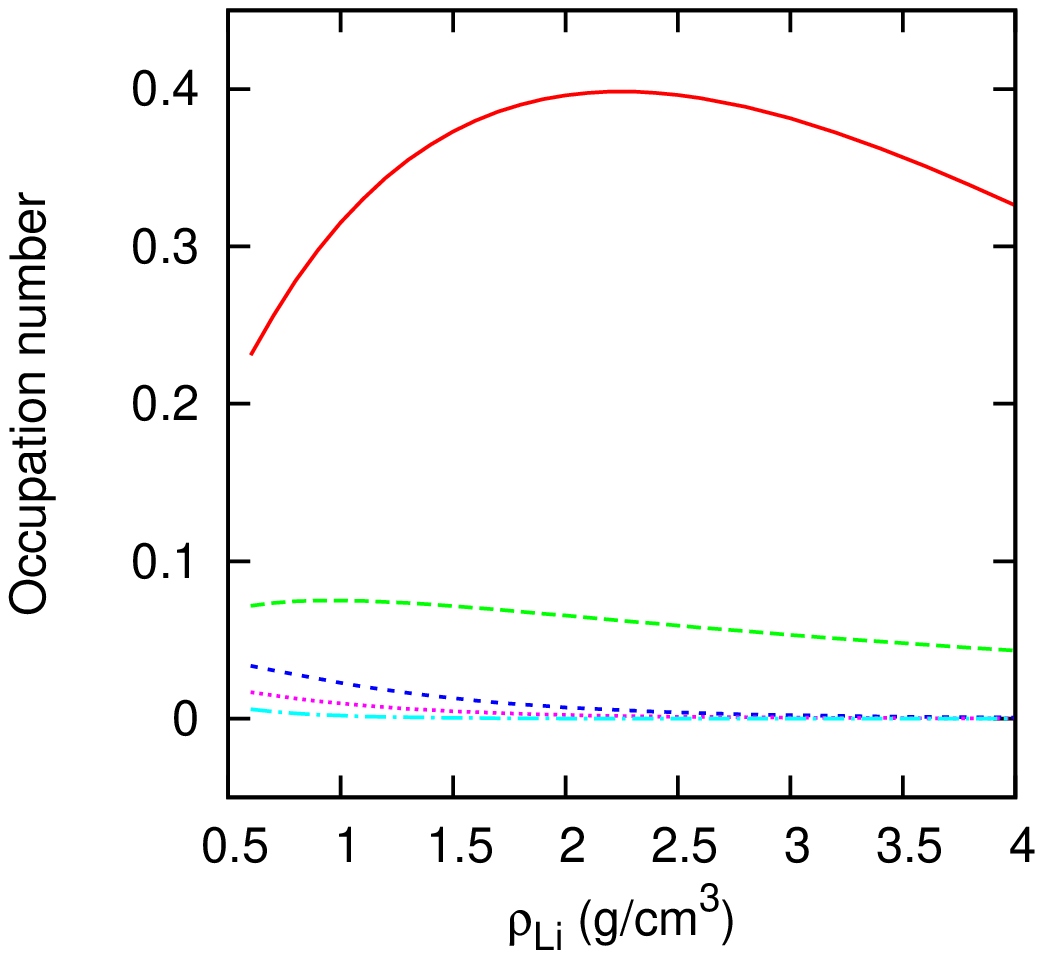}
\caption{Top: orbital energies for the $\Gamma$ point, 
at 100 kK.   Bottom left:
single-spin occupation number $f_i$ of the 
levels plotted for $\mathrm T=100$ K.  Bottom right: same but for $\mathrm T=100$ 
kK. The legend for the level number, given in the upper right, 
is for all plots.
}
\label{gammapoint}
\end{figure}

Again, we did calculations with ground-state XC on bcc Li, 
with material density from 0.6 to 4.0 g/cm$^3$. 
For all temperatures and densities, we used a 7 $\times$ 7 $\times$ 7
Monkhorst-Pack k-grid \cite{MonkhorstPack76}, a two-atom 
unit cell, and included 128 bands with
a plane wave energy cutoff of 150 Ry.  The calculations were done with
{\sc Quantum-Espresso} and the $1e^-$ PP just mentioned.  
The upper panel of Figure \ref{gammapoint} 
shows a sample of the orbital eigenvalues 
for a single k-point, $\Gamma$, as a function of
material density for ${\mathrm T} = 100$ kK.  The  ${\mathrm T} = 100$ K
plot is absolutely indistinguishable, due to relatively small differences 
in the eigenvalues.    
The eigenvalues are labeled in order of increasing energy, 
$\varepsilon_1$ lowest, $\varepsilon_{128}$  highest. 
The main point to be noticed is 
that as the density increases, the spread in the lower half (roughly) 
of the eigenvalues increases.  Those are the eigenvalues most
pertinent to
the calculation, in the sense that at a given temperature, 
excited levels will be depopulated at higher densities 
compared to the corresponding levels at lower densities.
(An exception would be a pressure-induced switch in level-ordering.) 
The lower panels of Fig.\ \ref{gammapoint} show the occupation  
numbers for those same eigenvalues. 
At low temperature and low density, the results are as expected, an
almost square-wave Fermi distribution with the lowest band fully 
occupied (since there are two electrons in the
unit cell) and the higher bands unoccupied. At higher
densities, some k-points, including the $\Gamma$ point, 
as shown in the lower left panel, for densities 
above 3 g/cm$^3$, have no occupation while others have two occupied
levels. This repopulation is a consequence of changes in the KS 
orbitals caused by changes 
in the external potential, hence also in the effective KS potential. The
lower right panel shows that at higher temperatures there is not only a
temperature dependence of the occupation numbers, but a significant
density dependence because of the spreading of the orbital energy
levels.

\begin{figure}
\includegraphics[angle=-90,width=8cm]{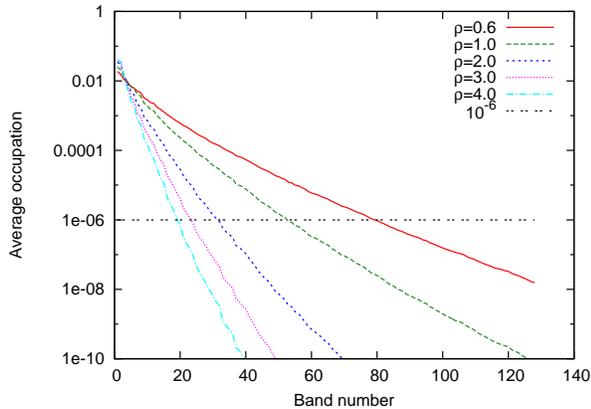}
\caption{
Zone-averaged band occupation numbers for all bands at $\mathrm T$ = 
100 kK, for the various material densities listed.
}
\label{bands}
\end{figure}

Next we consider the number of bands required for a stipulated
precision, given by a minimum occupation number threshold, as a function of 
temperature and density. For the calculations just discussed,  we
calculated a zone-averaged band occupation.   For a band of composite 
index $i$, we sum the occupations of the $\varepsilon_i$ level multiplied by 
the k-point integration weight for all $\mathbf k$-points.
This zone-averaged occupation is plotted for ${\mathrm T}=100$kK in 
Figure \ref{bands}.  One sees clearly that, 
for a given threshold in occupation number, for
example $10^{-6}$, the number of required bands decreases
significantly with increasing density. 
This decrease again is due to changes in the KS orbitals.  At least at
${\mathrm T}=0$K, it long has been known 
\cite{BoettgerTrickey85,ZittelEtAl85} that 
as Li is compressed, its band structure initially becomes less 
like the homogeneous electron gas (HEG) than the bcc zero-pressure 
bands.  However, eventually,
the system passes over to a Thomas-Fermi-Dirac equation of state,
signifying near-perfect but spread parabolic bands
(see upper panel of Fig. \ref{gammapoint}) and corresponding HEG 
occupations. 

\subsection{Exchange free energy}

Though it originated in the Greens function formalism of many-fermion
theory, the finite-temperature Hartree-Fock approximation is  
the thermodynamical generalization of
the variational optimization of a single-determinant trial
wave function which is ubiquitous in quantum-chemistry and molecular
physics as the HF approximation \cite{Mermin63}. 

To summarize, the thermal generalization of 
the familiar HF single-determinantal
exchange energy may be expressed in terms of the one-electron reduced 
density matrix (1-RDM)
%
\begin{align}
{\mathcal F}_{\rm x}[n] := - \int d{\mathbf x}_1 \, d{\mathbf x}_2 \, \lbrace &g_{12} \bar \Gamma^{(1)}({\mathbf x}_1|{\mathbf x}_2^{\prime})\nonumber \\ %
 &\times \bar\Gamma^{(1)}({\mathbf x}_2|{\mathbf x}_1^{\prime}) 
\rbrace_{{\mathbf x}_1^{\prime}={\mathbf x}_1, {\mathbf x}_2^{\prime}={\mathbf x}_2}\,,
\label{Ex}
\end{align}
%
where ${\mathbf x}:= {\mathbf r}, s$ is a composite space-spin variable, 
$g_{12} = 1/2{|{\mathbf r}_1-{\mathbf r}_2|}$, 
and the 1-RDM is defined in terms of the relevant orbitals 
$\{\varphi_{i}\}$ and occupation numbers $\{f_{i}\}$
\beq
 \bar \Gamma^{(1)}({\mathbf x}_1 |{\mathbf x}^{\prime}_{1})
 := \sum_{j=1}^{\infty} f_j \varphi_{j}({\mathbf x}_{1}) %
\varphi^{*}_{j}({\mathbf x}^{\prime}_{1})\,,
\label{1rdmA}
\eeq
subject to
\beq
f_j \equiv f(\varepsilon_j - \mu) = %
\lbrack 1 + \exp (\beta (\varepsilon_j - \mu))%
\rbrack^{-1}
\label{fsubj}
\eeq
and 
\bea
\int d{\mathbf x}  \varphi_{i}({\mathbf x}) %
\varphi^{*}_{j}({\mathbf x}) & = & \delta_{ij} \nonumber \\
 \sum_{j=1}^{\infty} f_j & = & N  \; ,
\label{1rdmB}
\eea
with $\beta := 1/k_B\rm T$ as usual.  Here $ \mu$ is the chemical potential 
(determined by Eq.\ (\ref{1rdmB})) 
and the $\varepsilon_j$ are the eigenvalues of the associated 
one-particle ftHF equation. 

The analogue to ftHF in 
DFT is called finite-temperature exact exchange  (ftEXX hereafter)
DFT \cite{Greiner10}.    In its pure Kohn-Sham form,
ftEXX defines the exchange free energy formally identically with 
ftHF, but evaluates the density $n({\mathbf r},\rm T)$ from orbitals
which follow from   
a true KS procedure, that is, from a one-body Hamiltonian 
with a local (multiplicative)  exchange potential.  That 
potential follows from the system response
function, $\delta n/\delta v_{\rm KS}$. A full ftDFT calculation (not
exchange only) would have a 
correlation free energy functional and associated KS potential as well.
  
Ground state DFT  with so-called hybrid approximate 
exchange functionals has a similar structure for the total energy, 
in the sense that hybrids have  
contributions both from single-determinant exchange and from 
exchange-correlation functionals which are explicitly density
dependent.  Instead of a KS procedure, one can go from 
such a hybrid expression directly to coupled one-electron equations 
by explicit variation with respect to the orbitals.  In ground-state
theory with a hybrid functional, this procedure sometimes is called 
generalized KS. 
The relevant point is that the same approach applies directly 
to ftHF.  Simply switch off the explicit density functionals
for exchange and correlation and leave the exchange functional
which comes from the trace over single-determinants.      
Since the capacity to do hybrid DFT 
as a generalized KS approach exists in both {\sc Vasp} and 
{\sc Quantum-Espresso}, one sees that such coding is immediately 
exploitable for doing ftHF.

In parallel with ground state DFT, an LDA may be obtained from
considering 
the finite-$\mathrm T$ HEG.  Its  
exchange free energy is given in first-order perturbation theory by
\begin{equation}
{\mathcal F}_{\rm x}^{\rm HEG} = -\frac{V}{(2\pi)^6}\int\int d\mathbf{k}\;d\mathbf{k^\prime} \frac{4\pi}{\left|\mathbf{k}-\mathbf{k^\prime}\right|}f({\mathbf{k}})f({\mathbf{k^\prime}})
\label{HEGfreeEn}
\end{equation}
where 
$f(\mathbf{k})=\lbrack 1 + \exp (\beta (k^2/2 - \mu^{\rm HEG}))\rbrack^{-1}$, 
and $V$ is the system volume. With the chemical potential expanded 
to the same order as well, 
$\mu^{\rm HEG} = \mu_0^{\rm HEG} +\mu_{\rm x}^{\rm HEG}$, the exchange portion 
is
\begin{equation}
\mu_{\rm x}^{\rm HEG}(n,{\rm T}) = \frac{\delta {\mathcal F}_{\rm x}^{\rm HEG}}{\delta n}
\label{muX}
\end{equation}
If expressed in closed form, this result may be used as the finite-$\mathrm T$
LDA, with exchange free energy per electron 
$f_{\rm x}^{\rm LDA}(n(\mathbf{r},{\rm T}),{\rm T}) =
({\mathcal F}_{\rm x}^{\rm HEG}/nV)|_{n=n({\mathbf r},T)}$,
 and 
$v_{\rm x}^{\rm LDA}(n({\bf r},{\rm T}),{\rm T})=
\mu_{\rm x}^{\rm HEG}(n,{\rm T})|_{n=n(\mathbf{r},{\rm T})}$.
Here 
we used the parametrization given by Perrot and Dharma-wardana 
\cite{Perrot.Dharma-wardana.1984}. The LDA exchange free energy
is then
\begin{equation}
  {\mathcal F}_{\rm x}^{\rm LDA}[n(\mathbf{r}),{\rm T}]=\int f_{\rm x}^{\rm LDA}(n(\mathbf{r},{\rm T}),{\rm T}) 
n(\mathbf{r},{\rm T}) d\mathbf{r}\,.
\label{tEx-LDA}
\end{equation}
The one-particle density follows by obvious analogy with
Eqs.\ \ref{1rdmA} - \ref{1rdmB}.

\subsection{Finite-temperature Hartree-Fock and DFT X-only calculations}

To study the importance of using an explicitly 
$\mathrm T$-dependent expression for the exchange free energy
(rather than a calculation with a ground-state X functional) 
and to estimate the quality of the $\mathrm T$-dependent exchange
free-energy functional defined by Eq.\ (\ref{tEx-LDA}),
we compare ftHF
calculations which use the exact exchange free energy Eq. (\ref{Ex}),  
Kohn-Sham calculations with $\mathrm T$-independent LDA exchange 
for the exchange free energy, 
${\mathcal F}_{\rm x} \approx E_{\rm x}^{\rm LDA}$, and 
KS calculations done with the $\mathrm T$-dependent exchange free energy 
functional 
${\mathcal F}_{\rm x}^{\rm LDA}$.  In the following discussion, the 
ground-state functional calculations are labeled ``LDAx'', while
those which used the explicitly $\mathrm T$-dependent LDA are labeled
``LDAx(T)''.  All the calculations were done with
{\sc Quantum-Espresso} using the $1e^-$ PZ LDA 
pseudopotential taken from the  {\sc Quantum-Espresso} web page.
We treated bcc Li with 
fixed nuclear positions, here with densities between $\rho_{\rm Li}=0.6$ and
$1.8$ g/cm$^3$ and temperatures between 100K and 100kK.  At these 
densities and temperatures, the {\sc Quantum-Espresso} 
$1e^-$ pseudopotential is adequate;
recall Sections  \ref{PAWhighdensLi} and \ref{FinTempPPlevel}  as
well as Figs.\ \ref{P-R.0.1kK.LSDA-abinit-vasp-siesta.R0.6-1.8} 
and 
\ref{P-T.LDA-TM-1e-3e.QEspresso}.

Convergence of the ftHF and
the LDAx calculations with respect 
to the ${\mathbf k}$-mesh for the bcc-Li 2-atom unit cell requires
attention.  It is known \cite{DacorognaCohen86} that T=0 K LDA 
calculations on bcc Li exhibit misleading convergence behavior 
at a relatively coarse $\mathbf k$-mesh density.   We tested 
for the smallest real space cell size used, 
corresponding to bulk density $\rho_{\rm Li}=1.8$ g/cm$^3$. 
The ftHF  total free energy calculations converge much more slowly 
than the ftDFT calculations.
For the moderate 
$7\times 7\times 7$ ${\mathbf k}$-mesh, the DFT calculations are 
converged to an iteration-to-iteration difference of 
0.02 eV per atom, while the HF calculations converge to the same
precision only upon reaching the much denser 
$17\times 17\times 17$ mesh.  Moreover, the HF calculation exhibits
a potentially misleading energy minimum at $15\times 15\times 15$.   
The ${\mathbf k}$-mesh convergence becomes faster with increasing $\mathrm T$.
For example,  at $\mathrm T$ = 100 kK, both HF and LDAx calculations
already are converged at the $3\times 3\times 3$ ${\mathbf k}$-mesh.  In
all calculations, both HF and DFT, presented in this section, 
the $25 \times 25 \times 25$ ${\mathbf k}$-mesh was used.

\begin{figure}
%
\includegraphics[angle=-00,height=3.3cm]{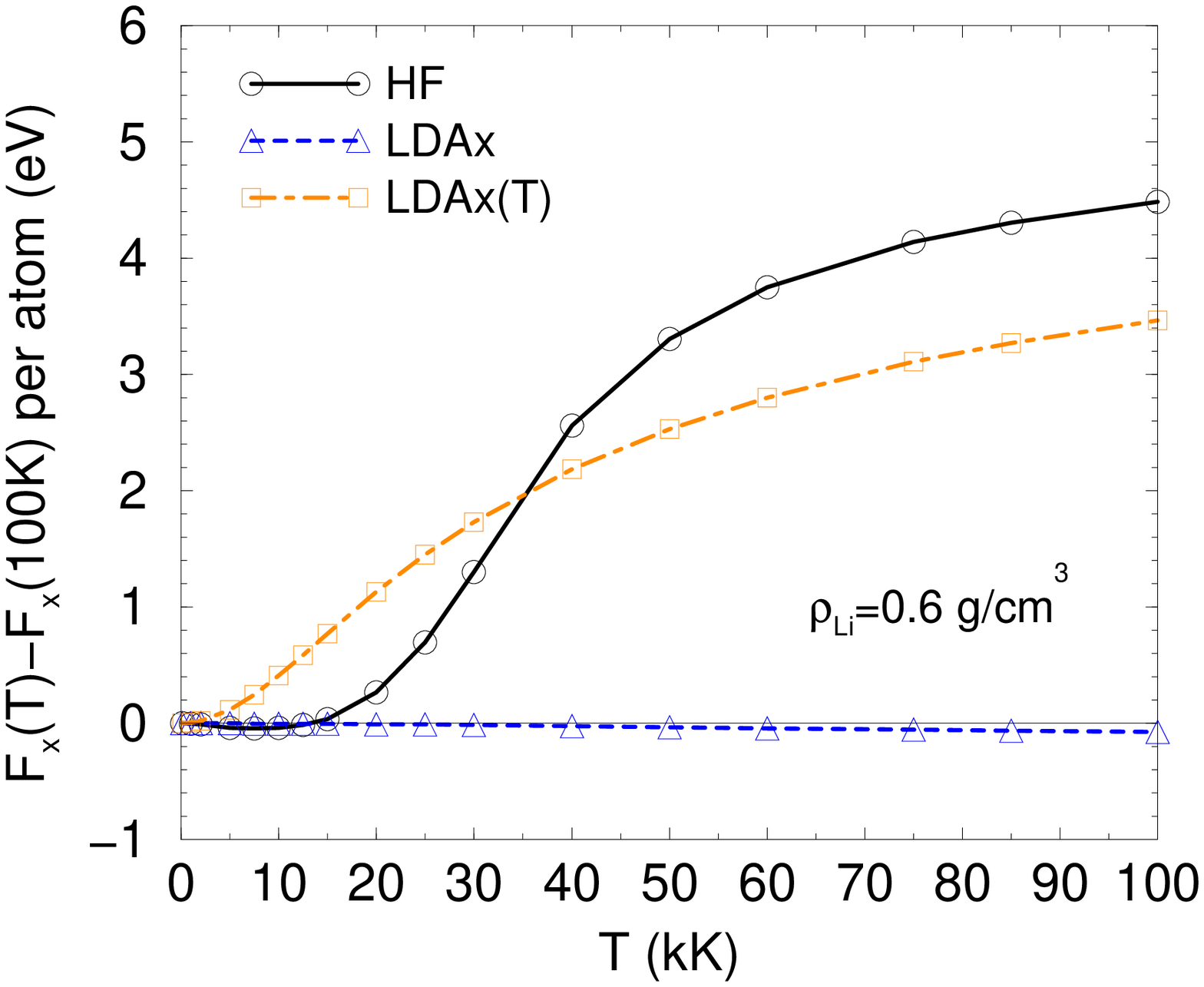}
\includegraphics[angle=-00,height=3.3cm]{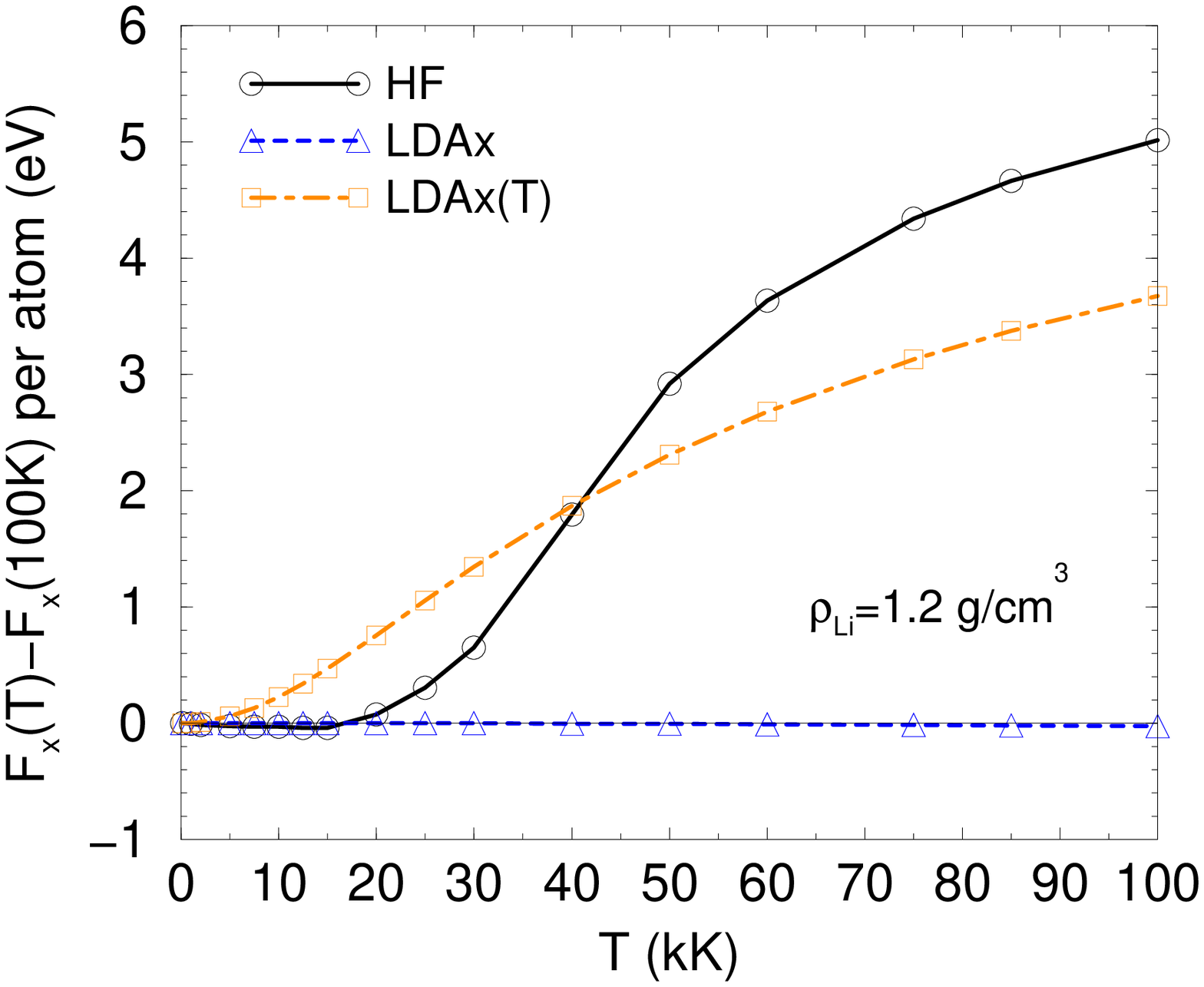}

\caption{
Comparison of finite-temperature HF, ground state LDA X-{\it only} (LDAx)
and $\mathrm T$-dependent LDA X-{\it only} (LDAx(T)) exchange free energy 
differences 
$\Delta {\mathcal F}_{\rm x}({\rm T})={\mathcal F}_{\rm x}({\rm T})-{\mathcal F}_{\rm x}(100 {\rm K})$ per atom as a function of 
electronic temperature $\mathrm T$.
Left panel: $\rho_{\rm Li}=0.6$ g/cm$^3$; right panel: $\rho_{\rm Li}=1.2$ g/cm$^3$.
}
\label{deltaEx-T.HFx-LDAx.QE}
\end{figure}

Figure \ref{deltaEx-T.HFx-LDAx.QE} compares changes in the exchange 
free energy contribution with increasing $\mathrm T$ relative to 100K values,  
${\mathcal F}_{\rm  x}({\rm T})-{\mathcal F}_{\rm x}(\rm 100 K)$. The 
$\mathrm T$-independent LDA
exchange free energy practically does not change over that range,
{\it i.e.}, ${\mathcal F}_{\rm x}^{\rm LDA}[n({\bf r},{\rm
  T})]\approx {\mathcal F}_{\rm x}^{\rm LDA}[n({\bf r},\rm 100 K)]$.  
In contrast, the HF  exchange free energy increases significantly 
(by about 4-5 eV per atom) with increasing $\mathrm T$. 
The $\mathrm T$-dependent LDA exchange free energy 
reproduces the  HF behavior at least qualitatively.  

\begin{figure}
\includegraphics[angle=-00,height=3.3cm]{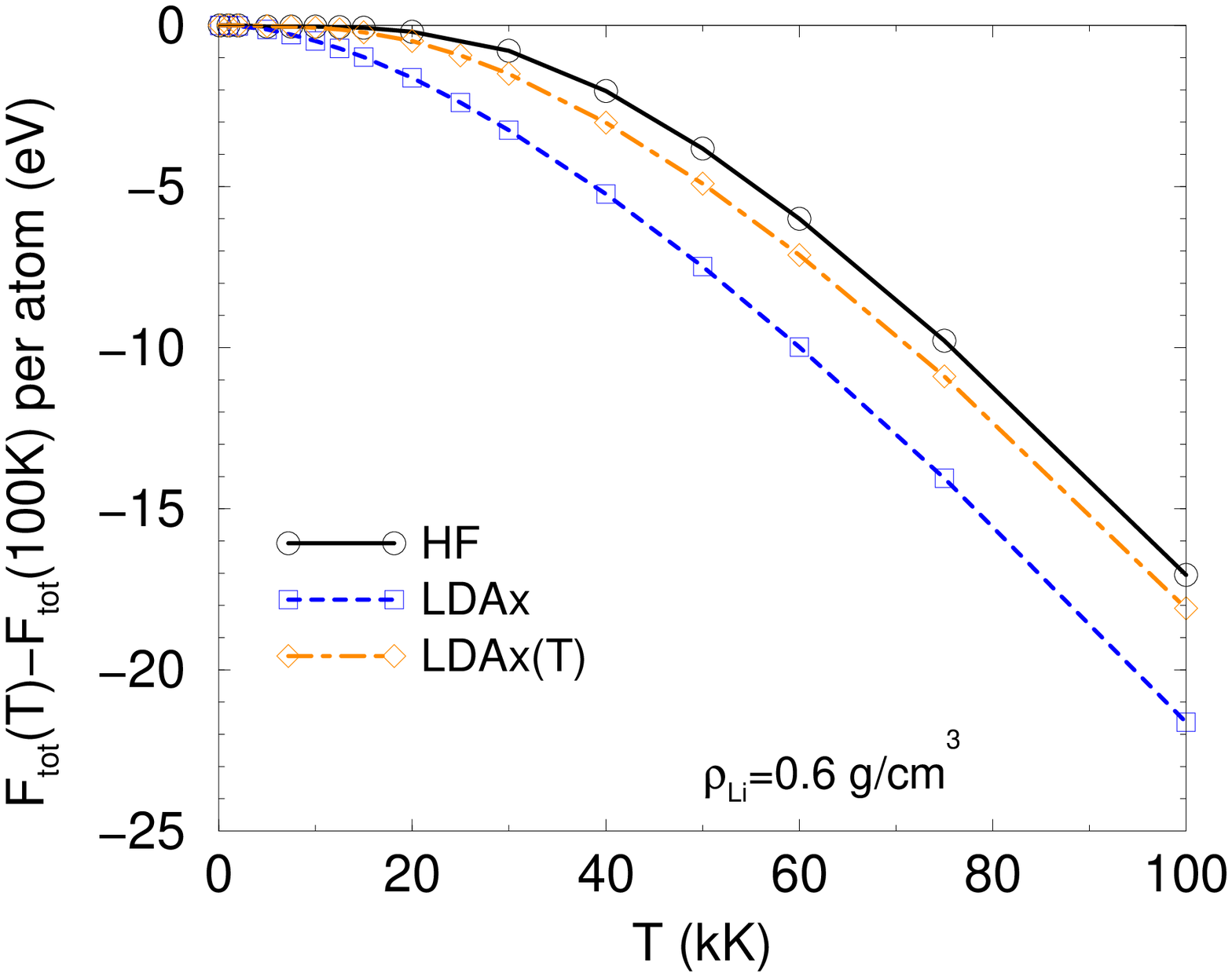}
\includegraphics[angle=-00,height=3.3cm]{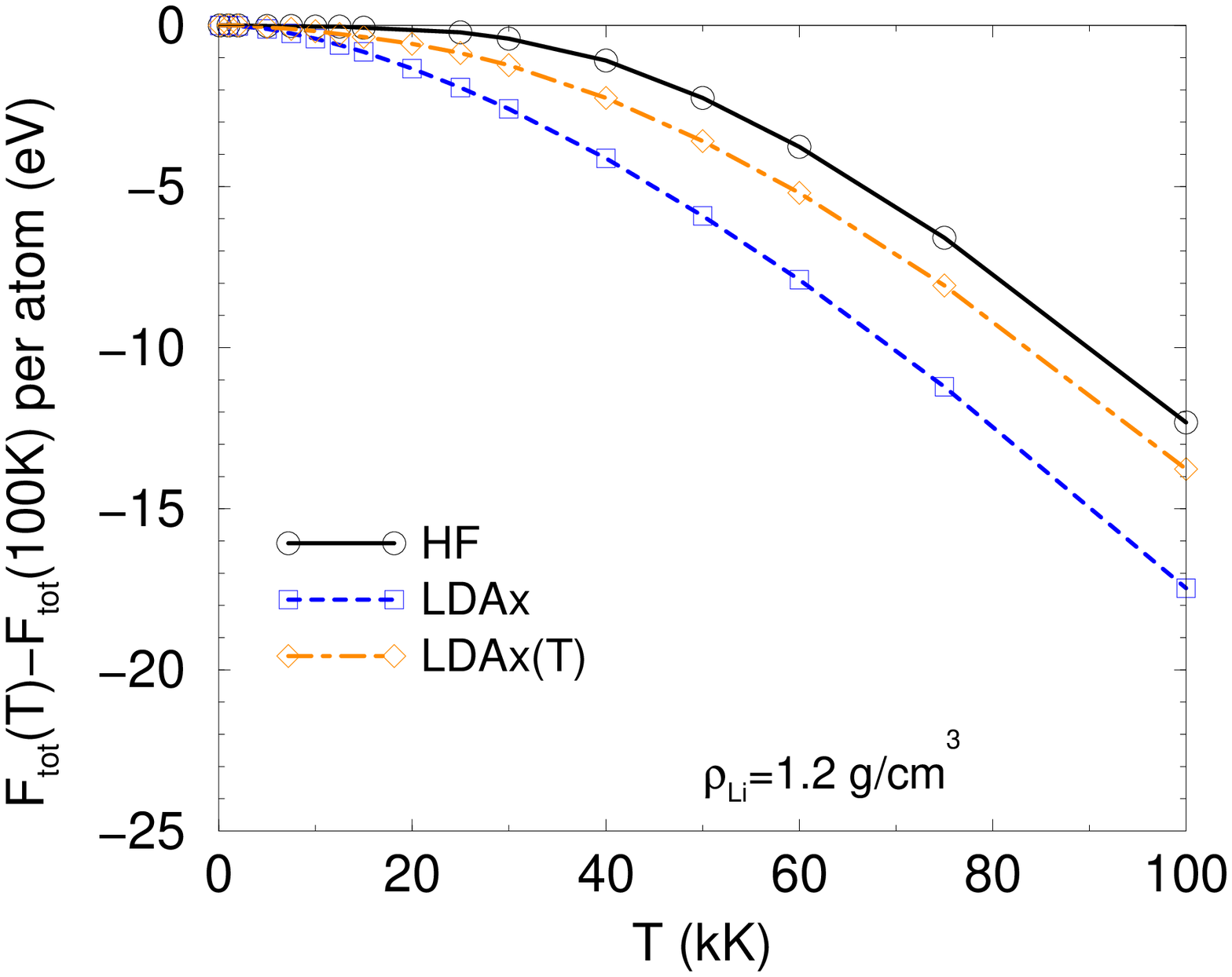}
\caption{
Comparison of finite-temperature HF, 
ground state LDA X-{\it only} (LDAx)
and $\mathrm T$-dependent LDA X-{\it only} (LDAx(T)) total 
free energy differences 
$\Delta {\mathcal F}_{\rm tot}({\rm T})={\mathcal F}_{\rm tot}({\rm T})-{\mathcal F}_{\rm tot}(100 {\rm K})$ 
per atom as a function of electronic temperature.
Left panel: $\rho_{\rm Li}=0.6$ g/cm$^3$; right panel: $\rho_{\rm Li}=1.2$ g/cm$^3$.
}
\label{deltaFtot-T.HFx-LDAx-TLDAx.QE}
\end{figure}

The exchange free energy is, of course, a small portion of the total
free energy. Figure \ref{deltaFtot-T.HFx-LDAx-TLDAx.QE} shows total
free energy differences $\Delta {\mathcal F}_{\rm tot}({\rm
  T})={\mathcal F}_{\rm tot}({\rm T})-{\mathcal F}_{\rm tot}(100 {\rm
  K})$ as a function of electronic temperature.  The free energy is
monotonically decreasing with increasing $\mathrm T$,  in
agreement with non-negativity of the entropy evaluated from the  thermodynamic
relation ${\mathcal S}=-\frac{\partial {\mathcal F}} {\partial \rm %
T}\Big|_{N,V}$.  The DFT X-{\it only} total free energies from 
$\mathrm T$-independent LDA X lie below the corresponding ftHF values
for all $\mathrm T$ and both densities.  At $\mathrm T \approx 20$ kK,
the interval is about 2 eV/atom, growing to about 4-5 eV/atom by 40 kK.  
The $\mathrm T$-dependent LDA X gives total free energy behavior much
closer to that of ftHF, with discrepancies not exceeding $1-2$ eV/atom.

\begin{figure}

\includegraphics[angle=-00,height=3.3cm]{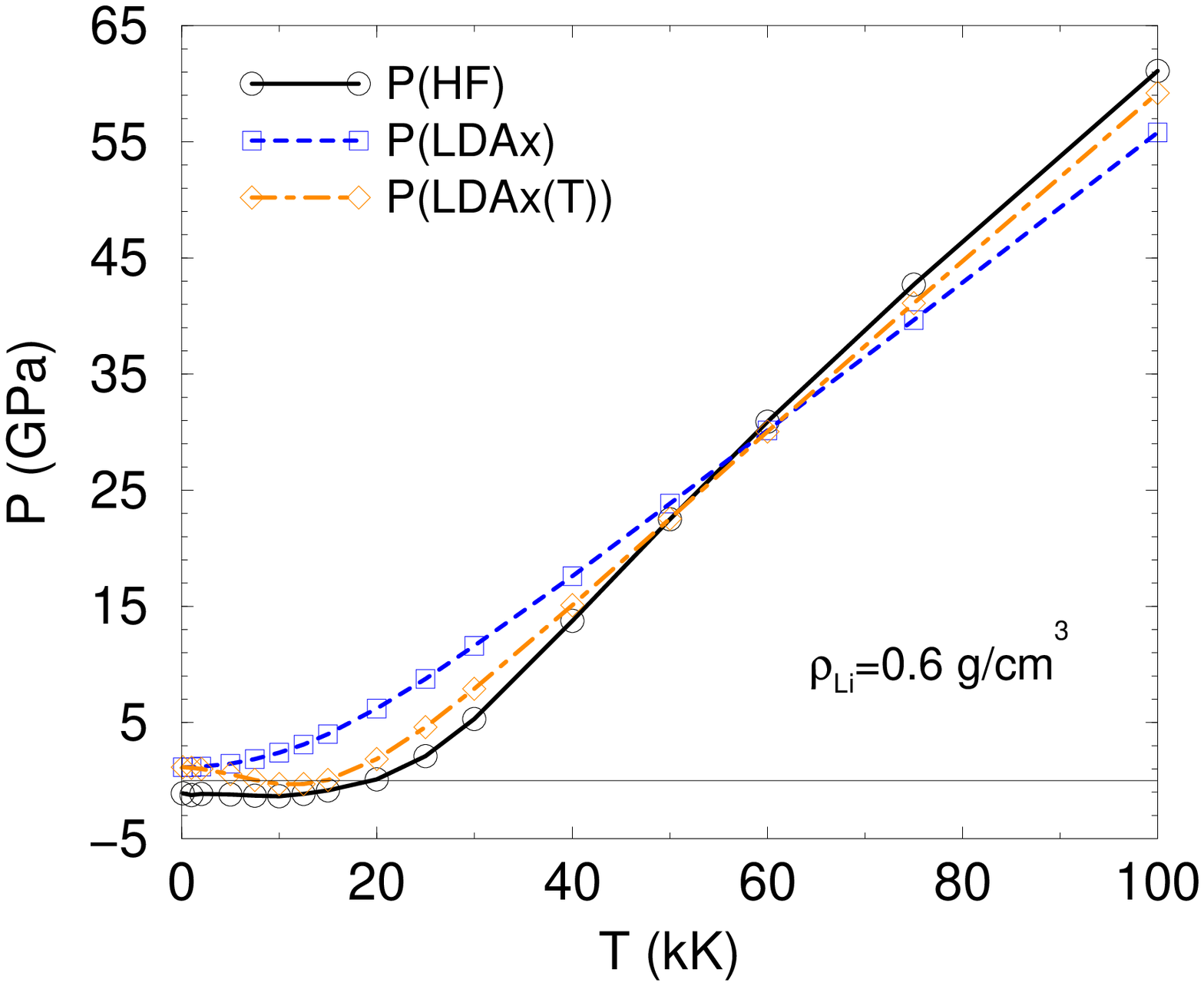}
\includegraphics[angle=-00,height=3.3cm]{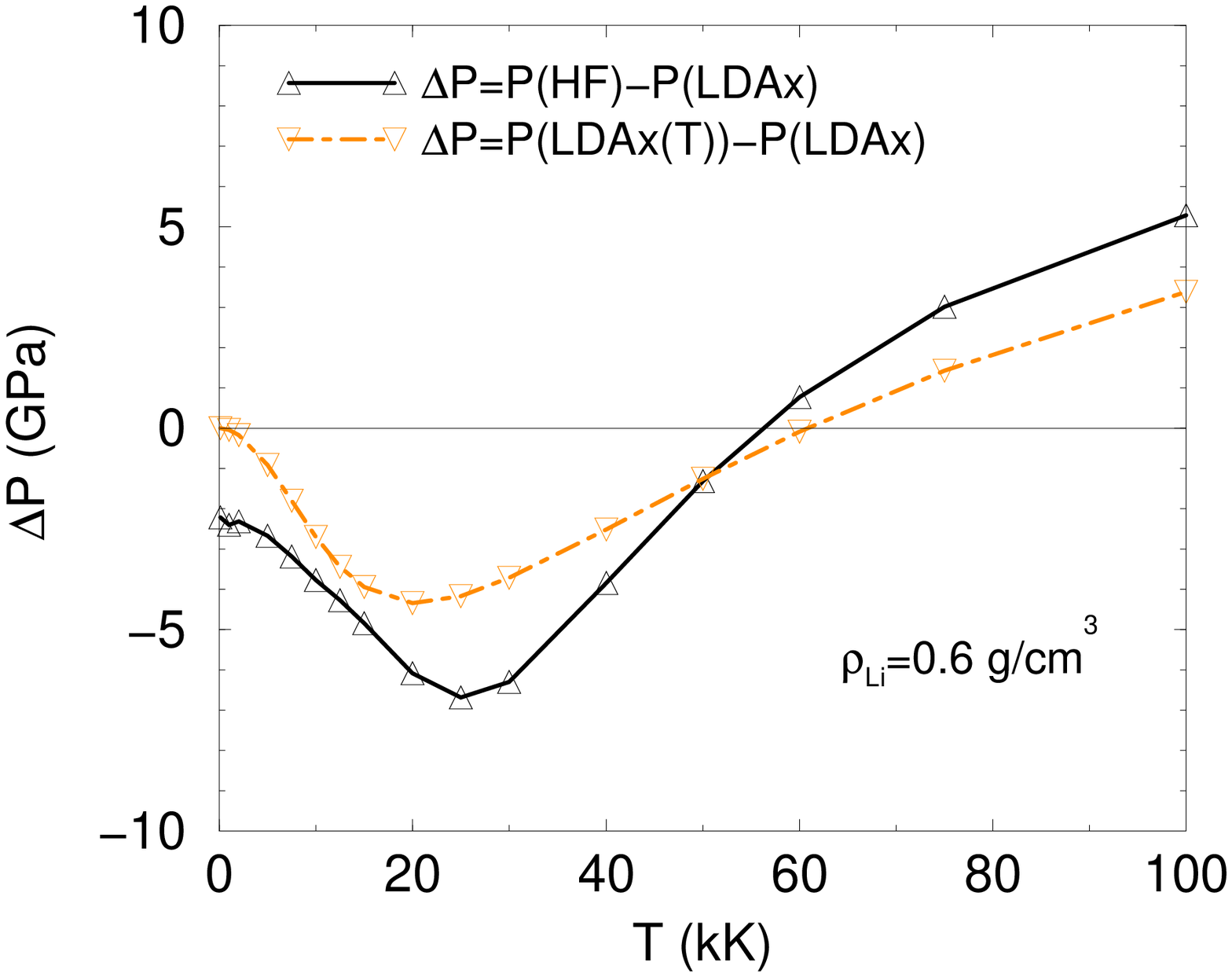}

\includegraphics[angle=-00,height=3.3cm]{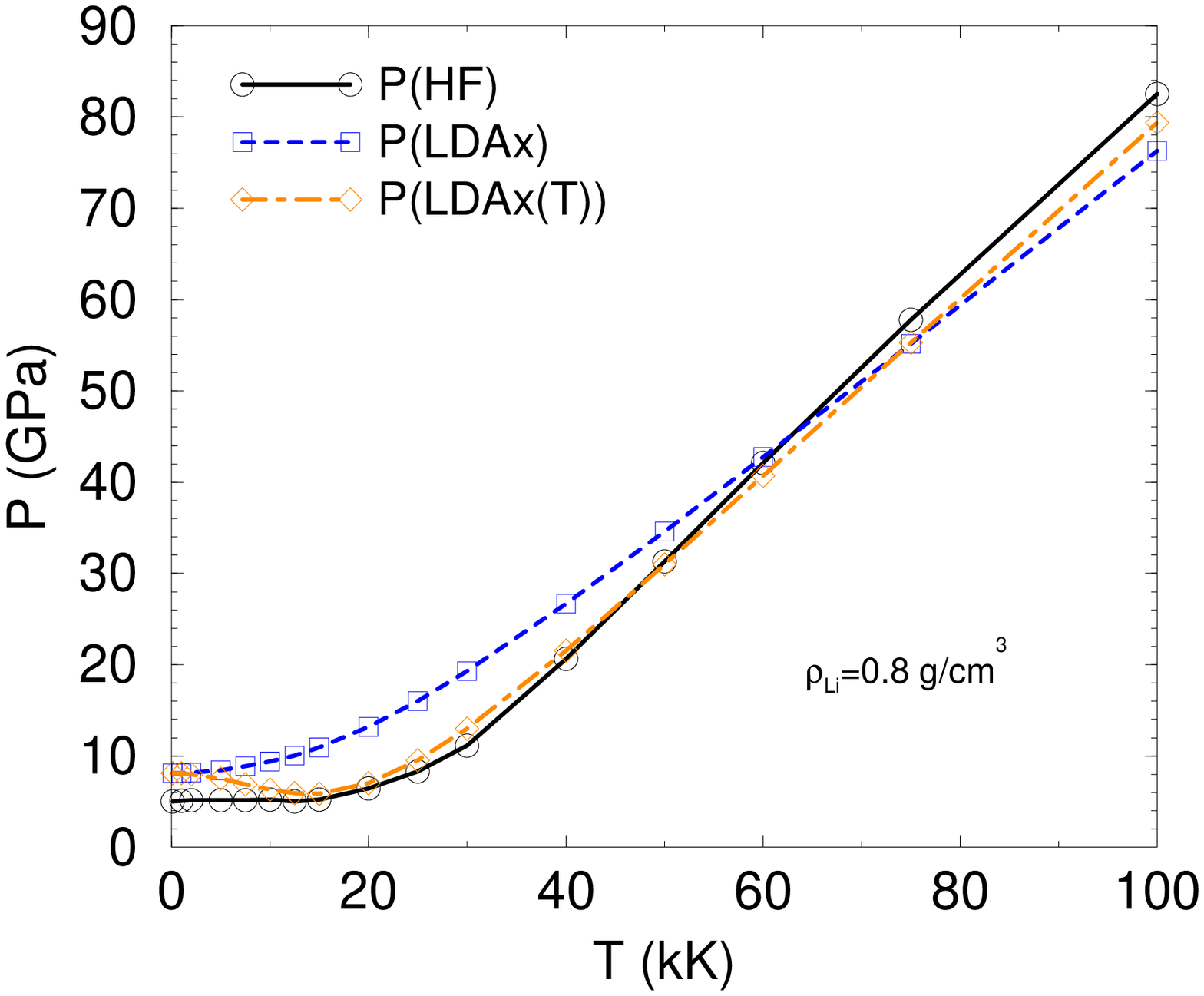}
\includegraphics[angle=-00,height=3.3cm]{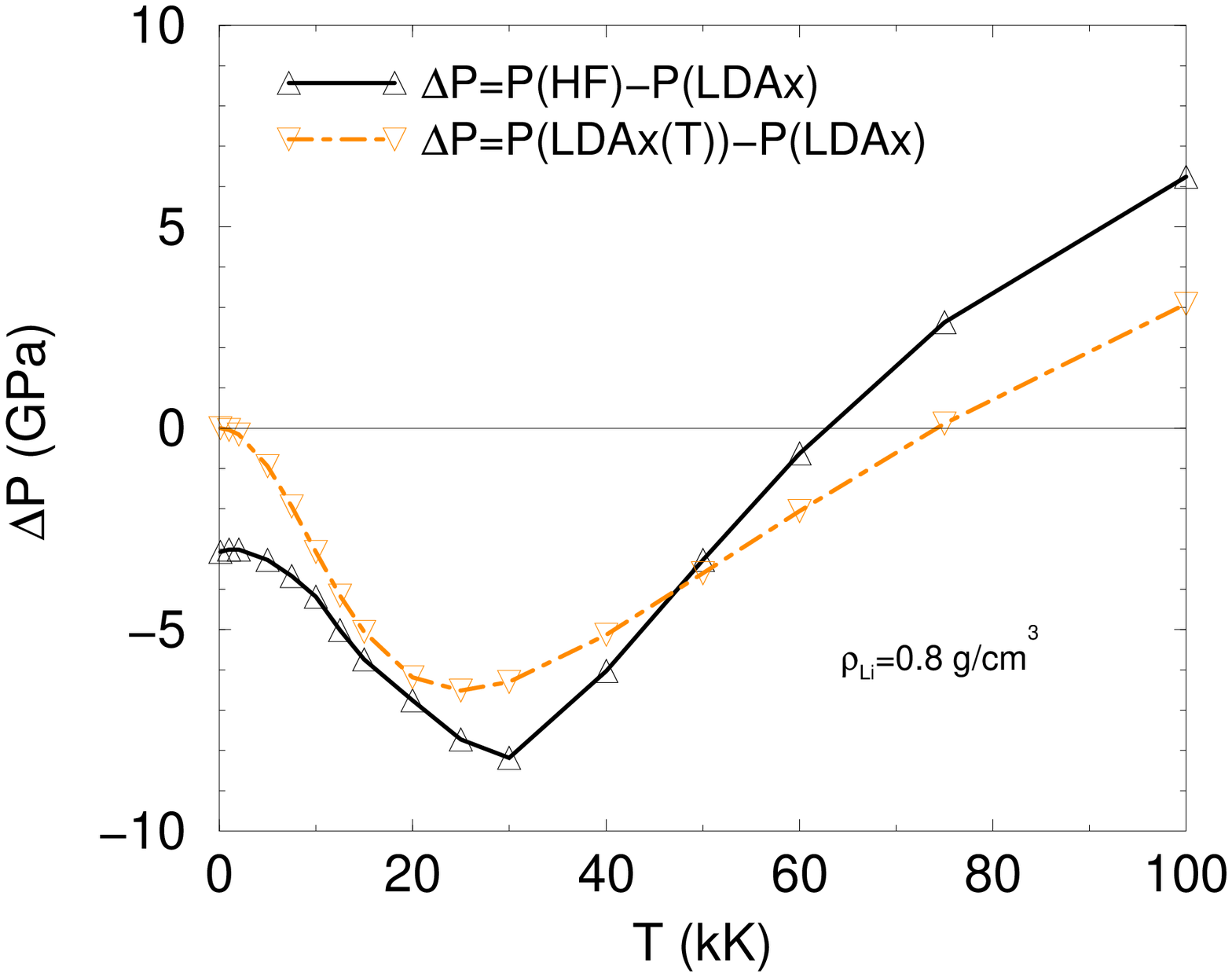}

\includegraphics[angle=-00,height=3.3cm]{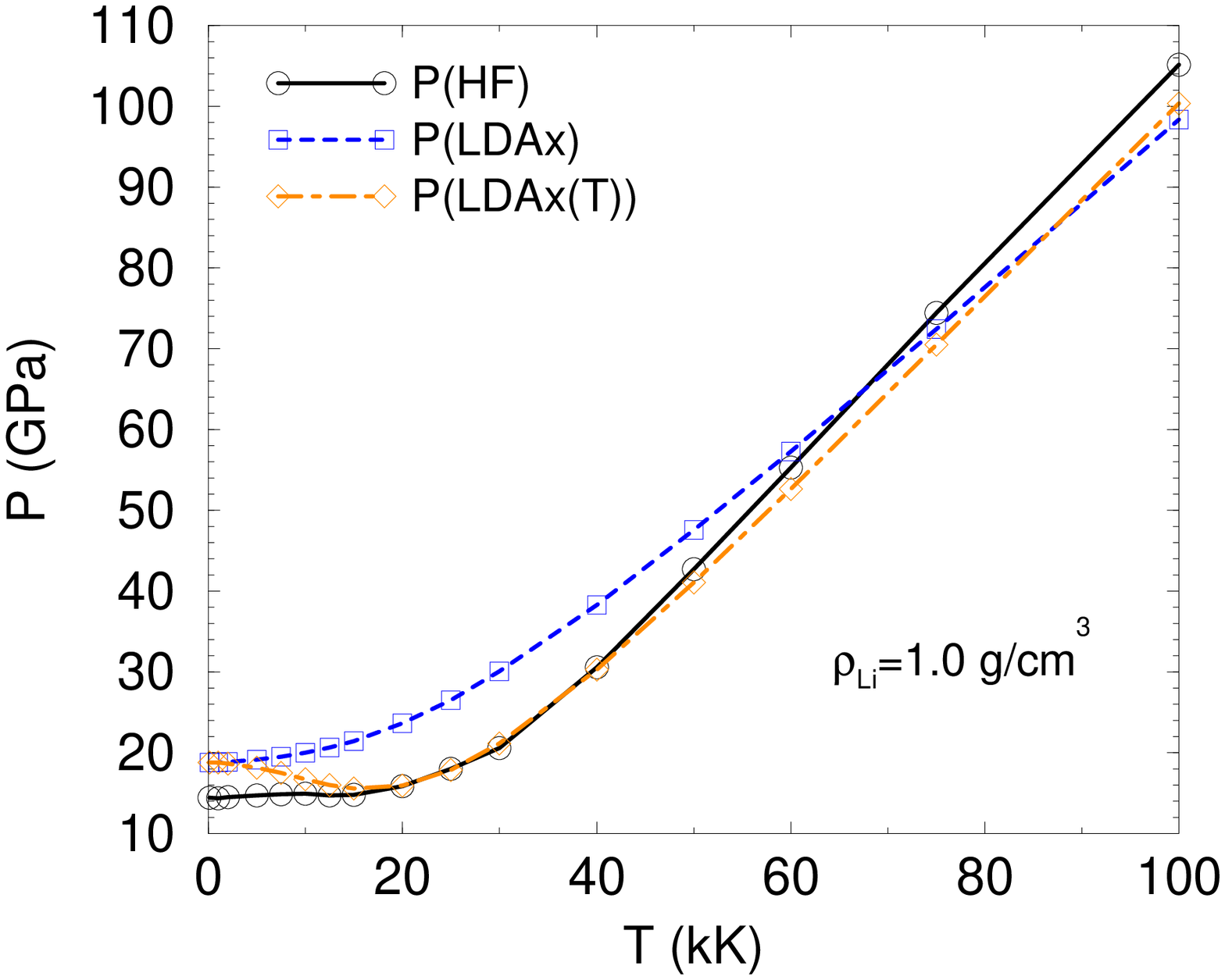}
\includegraphics[angle=-00,height=3.3cm]{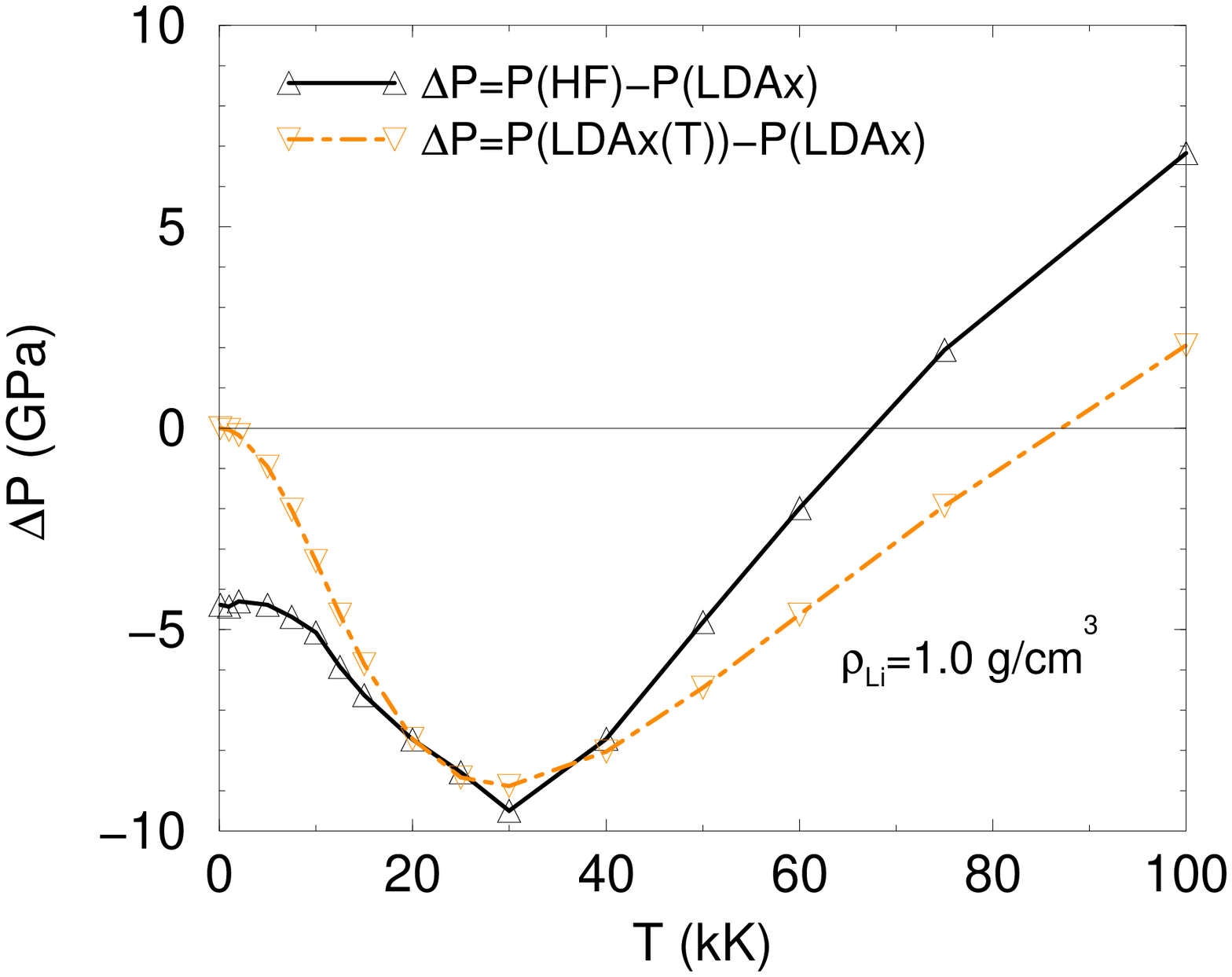}

\includegraphics[angle=-00,height=3.3cm]{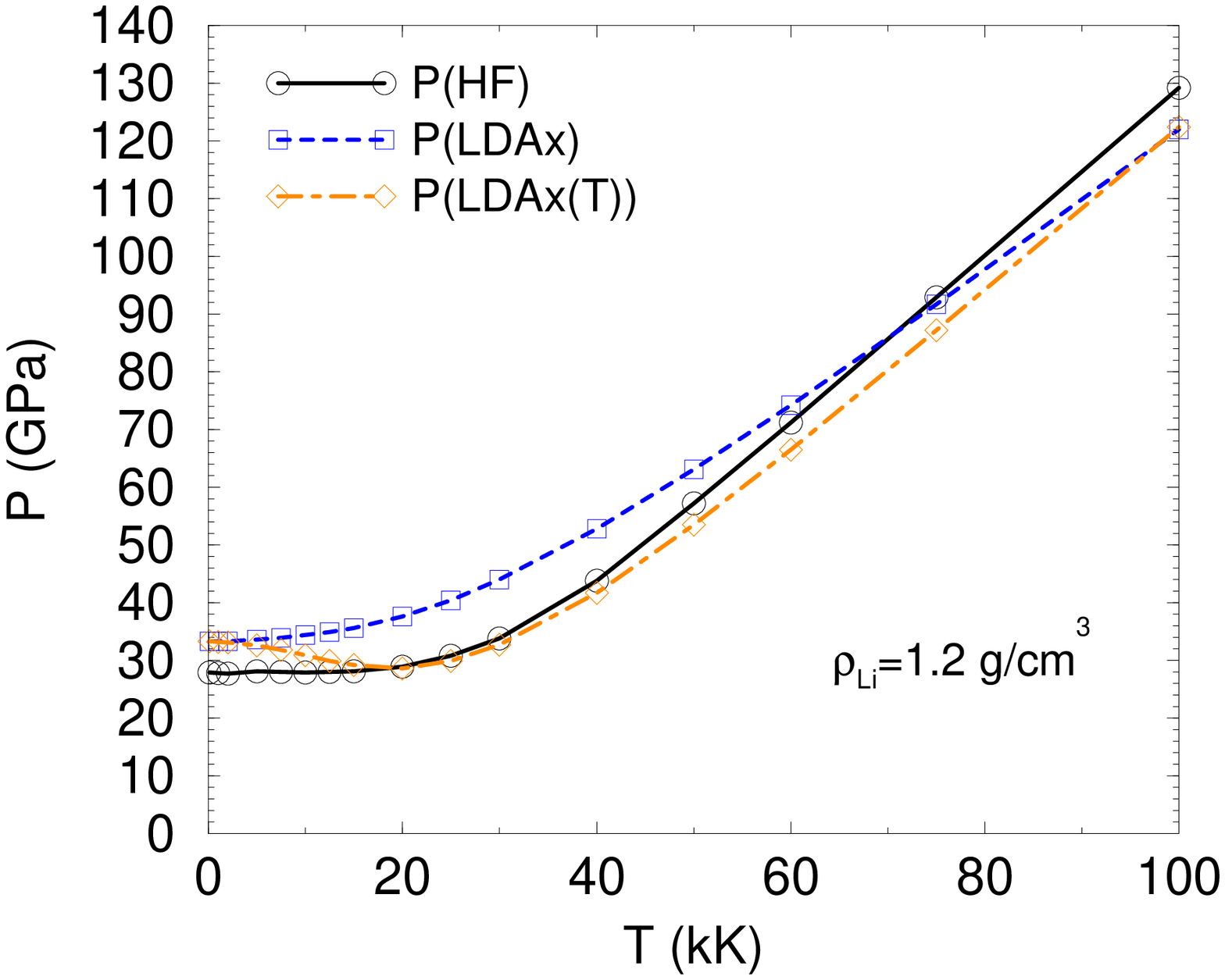}
\includegraphics[angle=-00,height=3.3cm]{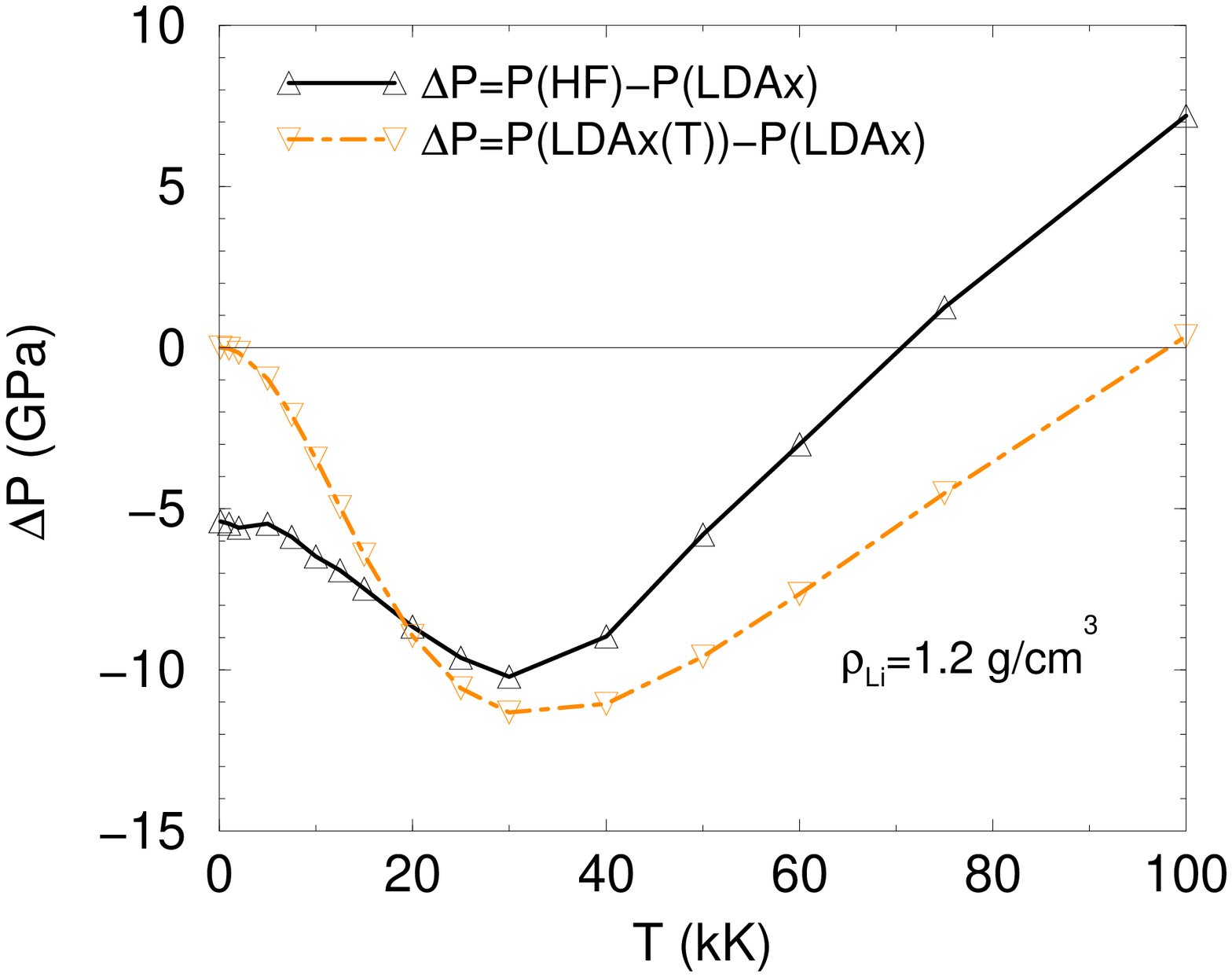}
\caption{
Effect of temperature-dependent exchange on pressure. 
Left panels: pressure as a function of electronic temperature as predicted
by HF, LDAx, and LDAx(T) calculations for $\rho_{\rm Li}=0.6, 0.8, 1.0$ 
and $1.2$ g/cm$^3$.
Right panels: differences in pressure between calculations with 
$\mathrm T$-dependent and $\mathrm T$-independent X, $\rm P(HF)-P(LDAx)$ and $\rm P(LDAx(T))-P(LDAx)$.
}
\vspace*{0.17cm}
\label{P-T.HFx-LDAx-TLDAx.QE}
\end{figure}

The effect of explicit $\mathrm T$-dependence in exchange upon the
pressure may be estimated from the difference 
between the ftHF or LDAx(T) and the LDAx values. 
Figure \ref{P-T.HFx-LDAx-TLDAx.QE} provides this comparison. 
As a function of $\mathrm T$, the pressure from ftHF  
starts below the LDAx curve, then crosses and 
goes above it at about  55-75 kK, depending upon the material density.
The temperature of this crossing point increases slightly with 
material density.  To isolate the effect of exact $\mathrm T$-dependent 
X upon the pressure, we consider the 
difference between ftHF and LDAx values, offset by the near-zero-temperature 
difference 
\be
\Delta P_{\rm HF-LDAx}({\rm T})=P_{\rm HF}({\rm T})-P_{\rm LDAx}({\rm T})
\ee
at ${\mathrm T} =$ 100K, {\it i.e.},
\bea
\Delta\Delta P_{\rm HF-LDAx}({\rm T})&=&\Delta P_{\rm HF-LDAx}({\rm T})%
\nonumber \\
&& - \Delta P_{\rm HF-LDAx}({\rm 100K})  \; .
\eea
One can see from the right-hand panels of Fig.\ \ref{P-T.HFx-LDAx-TLDAx.QE}
that the maximum magnitude of this difference at $\mathrm T$= 100 kK is about 
10 \% for all material densities considered.   For low temperatures, the 
effect of exact $\mathrm T$-dependent X on pressure
is stronger.  For example, at 30 kK
$\Delta\Delta P_{\rm HF-LDAx}({\rm 30 kK})\approx 5$ GPa  
for material density 1.0 g/cm$^3$, the shift is 
$\approx$ 30\% of the HF pressure (about 15 GPa) at that $\mathrm T$.
Again, the LDAx(T) and ftHF temperature-dependence resemble one another
qualitatively whereas the LDAx result does not.  Note that the 
LDAx(T) crossing temperature 
with respect to the LDAx curve increases much more rapidly with 
increasing material density than for ftHF. For material 
density $\rho_{\rm Li}=0.6$ g/cm$^3$, both curves 
cross at ${\mathrm T} \approx 60$ kK. 
At $\rho_{\rm Li}= 1.2$ g/cm$^3$ the LDAx(T) pressures crosses the LDAx curve 
at ${\mathrm T}\approx$ 100 kK,  higher than the temperature of the HF-LDAx 
crossing point ${\mathrm T} \approx$ 70 kK. With increasing material 
density the shift between LDAx(T) and ftHF increases especially 
for ${\mathrm T} \ge$ 50 kK.

\section{Conclusions}

Detailed computational examination of the applicability of standard PP 
and PAW methods to the WDM regime,  
with bulk Li as the test system, yields several insights.  
By unambiguous comparison with all-electron results from 
small Li clusters of bcc-derived symmetry, we find that the PAW 
scheme requires a small augmentation sphere radius, that the 
compensation-charge term is not helpful, and that all electrons 
must be treated in the SCF calculation.   
We have constructed such PAW data sets for LDA and GGA functionals and 
used them to generate reference data. 

We have located the maximal material density of bulk
bcc-Li usable for standard PPs in {\sc Vasp}, {\sc Abinit}, and 
{\sc  Quantum-Espresso} codes.  And we have delineated the validity of
using such PPs at high $\mathrm T$ by  comparison of
$1e^-$ and $3e^-$ PP results.  
The transferability of PPs and PAW data sets
developed for near-equilibrium conditions to the WDM regime is 
conditional.  At near-equilibrium densities it appears to be acceptable,
but not at high densities.  Clearly, such transferability should not be
assumed.  

With these issues settled,  we have found that there is non-trivial
effect of explicit $\mathrm T$-dependence in the X functional in 
the specific sense of   
comparison with ftHF.  In particular, 
the LDA $\mathrm T$-dependent exchange contribution to the total
free energy is much closer to the exact HF exchange value than 
is the contribution from exchange approximated by
the LDA ground-state X functional.
Although the
exchange free energy is a small portion of the total free energy, this
difference carries over into clearly significant differences in the
equation of state.  Thus, the 
effect of explicit $\mathrm T$-dependence in X 
is relevant for an accurate characterization of the Li equation of state 
in the WDM regime.   We suspect that this may be generally true of
WDM systems.  If so, 
$\mathrm T$-dependent LDA exchange  may serve 
as a starting point for development of more refined GGA-type
exchange free energy functionals, analogous with the role of LDA
in the ground state.

\section{Acknowledgments}
We acknowledge, with thanks, many informative conversations with
our colleagues J.W.\ Dufty, F.E.\ Harris, and K.\ Runge. 
SBT also thanks 
M.\ Desjarlais, F.\ Lambert, and L.\ Collins for informative
discussions.
Work supported in part under U.S.\ Dept.\ of Energy BES (TCMP, TMS) grant
DE-SC 0002139.

\end{document}